\documentclass[aps,prl,showpacs,amsmath,reprint,superscriptaddress,floatfix,nobalancelastpage,footinbib]{revtex4-2}

\usepackage[colorlinks=true, citecolor=black, linkcolor=black, urlcolor=black]{hyperref}
\usepackage[all]{hypcap}
\usepackage{amsmath}
\usepackage{enumerate,bm}
\usepackage{graphicx}
\usepackage{color}
\usepackage{esint}
\usepackage[usenames,dvipsnames]{xcolor}
\usepackage{subfigure}
\usepackage{flushend}
\usepackage{amssymb}
\usepackage{float}
\usepackage{dcolumn}
\usepackage{ulem} 

\usepackage{setspace}
\usepackage{verbatim}

\begin{document}

\author{Dongbin~Shin}
\email{dshin@gist.ac.kr}
\affiliation{Department of Physics and Photon Science, Gwangju Institute of Science and Technology (GIST), Gwangju 61005, Republic of Korea}
\affiliation{Max Planck Institute for the Structure and Dynamics of Matter and Center for Free Electron Laser Science, 22761 Hamburg, Germany}

\author{Angel~Rubio}
\email{angel.rubio@mpsd.mpg.de}
\affiliation{Max Planck Institute for the Structure and Dynamics of Matter and Center for Free Electron Laser Science, 22761 Hamburg, Germany}
\affiliation{Nano-Bio Spectroscopy Group, Departamento de Fisica de Materiales, Universidad del País Vasco UPV/EHU- 20018 San Sebastián, Spain}
\affiliation{Center for Computational Quantum Physics (CCQ), The Flatiron Institute, 162 Fifth avenue, New York NY 10010.}

\author{Peizhe~Tang}
\email{peizhet@buaa.edu.cn}
\affiliation{School of Materials Science and Engineering, Beihang University, Beijing 100191, P. R. China}
\affiliation{Max Planck Institute for the Structure and Dynamics of Matter and Center for Free Electron Laser Science, 22761 Hamburg, Germany}

\title{Light-induced ideal Weyl semimetal in HgTe via nonlinear phononics}


\date{\today}

\begin{abstract} 
Interactions between light and matter allow the realization of out-of-equilibrium states in quantum solids.
In particular, nonlinear phononics is one of the efficient approaches to realizing the stationary electronic state in non-equilibrium.
Herein, by using extended $ab~initio$ molecular dynamics, we identify that long-lived light-driven quasi-stationary geometry could stabilize the topological nature in the material family of HgTe compounds.
We show that coherent excitation of the infrared-active phonon mode results in a distortion of the atomic geometry with a lifetime of several picoseconds. 
We show that four Weyl points are located exactly at the Fermi level in this non-equilibrium geometry, making it an ideal long-lived metastable Weyl semimetal. 
We propose that such a metastable topological phase can be identified by photoelectron spectroscopy of the Fermi arc surface states or ultrafast pump-probe transport measurements of the nonlinear Hall effect.
\end{abstract}

\maketitle

Non-equilibrium states driven by strong light-matter interactions have attracted a lot of attention in recent years, due to intriguing physical phenomena without equilibrium counterparts and interesting ultrafast applications~\cite{Terro2021,dante2022,Bao2022,disa2021engineering,Basov2020,Bloch2022}.
Unlike modulating ground states, the coupling between the laser field and matter often produces non-thermal electronic/phononic changes in their electronic structures and charge populations.
For example, Floquet engineering is expected to be realized by applying a continuous light~\cite{oka2018,Bao2022,Hannes2021,Rugg2018,disa2021engineering,Bloch2022,Basov2020,zhou2023pseudospin,ZhouBP2023_PRL}. 
Recent theoretical works have suggested that coupling between a time-periodic oscillating field and the Dirac bands could result in a topologically non-trivial insulating state among the Floquet spectra~\cite{rudner2020band}.
However, the electronic Floquet states are difficult to be achieved and only reported on a few materials \cite{Wang2013,zhou2023pseudospin,GierzWSe2} because of dissipation, decoherence, and heating effects~\cite{Iadecola2015,GierzWSe2}.
In contrast, nonlinear coupling between materials and phonon via terahertz (THz) pulses can lead to robust quasi-stationary states~\cite{disa2021engineering,Shin2020,forst2011nonlinear,Mitrano2016,Buzzi2020,Kozina2019,Subedi2015}, such an approach is named the nonlinear phononics. We consider an infrared (IR) active phonon mode resonantly excited by the THz pulse, its oscillation amplitude can be largely enhanced to an anharmonic range. Via the nonlinear phonon interaction coupled with another phonon (e.g. $R_{c}$ mode), a time-averaged atomic displacement can be induced along $R_{c}$'s oscillating direction, resulting in an effective lattice distortion in ultrafast time scale. Consequently, electronic properties can be efficiently manipulated, realizing long-lived non-equilibrium states~\cite{Mankowsky2016,Basov2017}. In this context, light-induced superconductivity has been experimentally shown in cuprate, alkali-doped C$_{60}$, and organic conductors~\cite{Fausti2011,Mitrano2016,Buzzi2020}. Light-induced magnetic orderings in perovskite ErFeO$_3$ and piezomagnetic crystal CoF$_2$~\cite{Nova2017,Disa2020}, as well as a light-induced ferroelectric transition in SrTiO$_3$~\cite{Kozina2019,Li2019,Nova2019,Shin2021arxiv,Simone2021} have been assessed from the same perspective.
On the theoretical side, the light-induced non-equilibrium states induced by nonlinear phononics have usually been investigated using a model Hamiltonian that includes the relevant coupling to specific modes and some phenomenological dissipative terms~\cite{Kozina2019,Shin2021arxiv,Subedi2015,pueyo2022,subedi2014}.
In this regard, first-principles calculations considering non-equilibrium dynamics, including both couplings to the external field and realistic dissipation channels, are needed. 

As a notable exotic quantum state of matter, topological Dirac and Weyl semimetals host gapless excitations with topological charges in the bulk and Fermi arc states on the surface~\cite{Armitage2018,liu2020semimetals}, which have distinctive electronic and optical responses. The electronic structures of these semimetals could be affected by light-matter interactions. 
For example, the light-induced anomalous Hall effect has been observed in the graphene~\cite{McIver2020}, where the asymmetric photo-carrier population effect induced by the circularly polarized light contributes to an anomalous Hall signal~\cite{Sato2019}.
Furthermore, the phase transition from a Dirac semimetal to a Floquet Weyl semimetal in NaBi$_3$ was signified as an intentional breaking of the time-reversal symmetry by the driven light field~\cite{hubener2017creating}.
On the other hand, a long-lived meta-stable lattice distortion could also result in the stationary states in non-equilibrium~\cite{Sie2019,Vaswani2020}.
For example, the driven phase transition is observed in bulk WTe$_2$~\cite{Sie2019,guan2021manipulating} and ZrTe$_5$~\cite{Luo2021}. Experimental evidence for photo-carrier-induced non-equilibrium dynamics in Weyl materials are reported for $\beta$-WP$_2$~\cite{Hu2023} and elemental tellurium~\cite{Ning2022,Jnawali2020} recently, in which the possible topological phase transition and topological band modulations have been observed.
However, because of various excrescent contributions from non-topological states near the Fermi level in these materials mentioned above~\cite{Deng2016,Armitage2018}, the ultrafast transport and optical signals will be strongly influenced by these trivial states and cannot be entirely ascribed to light-induced topological characteristics.

Herein, we demonstrate a robust metastable light-induced topological phase transition to an ideal Weyl semimetal from a trivial semimetallic state in the family of bulk HgTe, as shown in the schema in Figs. 1(a) and 1(b).
To achieve this, we have employed a new implementation of $ab~initio$ molecular dynamics (AIMD), which additionally introduces the direct interaction between the classical ions and the oscillating external electric field, including realistic nonlinear phonon interactions and dissipation via phonon-phonon scatterings~\cite{Kozina2019,Shin2021arxiv,Subedi2015,Lu2022,Malic2011}.
In this way, we identify a non-equilibrium geometry with distortion along the combined direction of IR modes (named as IR$_{y}$ and IR$_{z}$ shown Figs.~1(d) and~1(e)) when the IR$_{x}$ mode in HgTe is resonantly excited (see Fig. 1(c)). 
The new steady structure of HgTe hosts an ideal Weyl semimetal phase with a lifetime of several picoseconds. 
When we increase the laser intensity, the amplitude of the atomic distortion is enhanced, further enlarging the distance between Weyl points in bulk and the length of the Fermi arc on the surface. 
In addition, similar THz field-induced dynamics are observed in bulk HgSe and $\beta$-HgS.
Because the light-driven HgTe is an ideal Weyl semimetal without the trivial band contributions around the Fermi level, the signatures measured by time- and angle-resolved photoemission spectroscopy (TrARPES) and nonlinear Hall conductivity~\cite{Xu2015,Yang2015,Gierz2013} have a clear topological origin.

\begin{figure}[t]
  \centering
  {\includegraphics[width=0.35\textwidth]{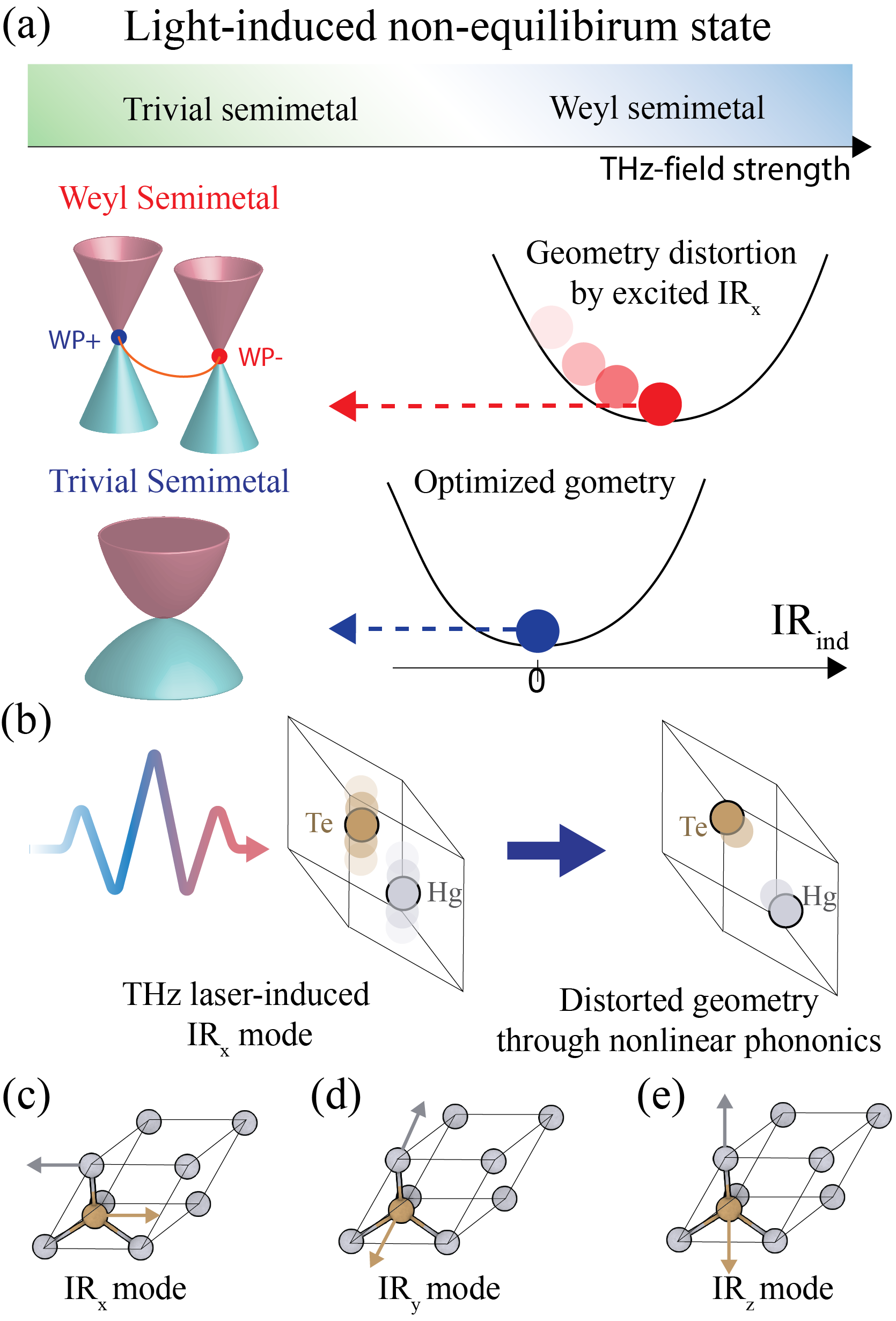}}
   \caption{
    Light-induced Weyl points in bulk HgTe through the nonlinear phonon interaction between IR modes.
   (a) Schematic phase diagram of THz field-induced Weyl points from trivial semi-metal HgTe and the modification of potential energy surface for induced phonon mode under the excited IR$_x$ mode.
   (b) Schematic image of THz field-induced atomic geometry distortion of HgTe through the nonlinear phonon interaction.
   Three degenerated IR phonon modes along (c) x-direction, (d) y-direction, and (e) z-direction.
}\label{Fig1}
\end{figure}

\begin{figure}[t]
  \centering
  {\includegraphics[width=0.48\textwidth]{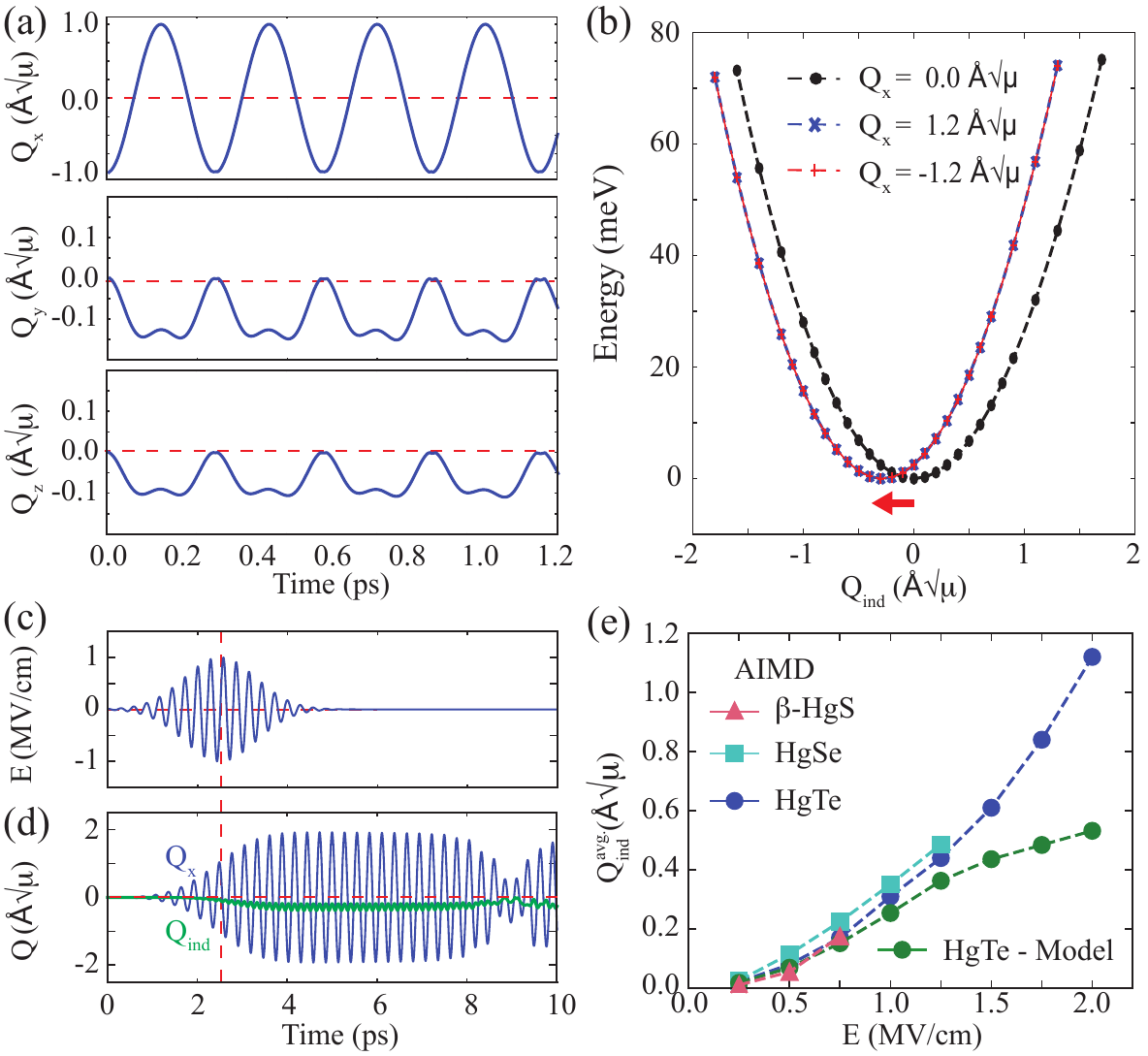}}
   \caption{
   Nonlinear phonon interaction in HgTe and light-induced atomic geometry distortion.
   (a) Nonlinear phonon dynamics of IR phonon modes simulated by AIMD simulation.
   (b) Modified potential energy surface of $Q_{ind}$ by shifted atomic geometry toward $Q_x$.
   Time profile of (c) THz field and (d) induced phonon modes evaluated from AIMD simulations.
   (e) E-field strength dependence of geometry distortion along $Q_{ind}$ in HgTe family.
}\label{Fig2}
\end{figure}

Before discussing the light-induced dynamical simulations, we describe the vibrational properties and nonlinear phonon interactions in HgTe.
In ambient conditions, bulk HgTe with the Zinc Blende structure hosts three degenerate IR phonon modes at the $\Gamma$ point (see Figs. 1(c)-1(e)).
To investigate their nonlinear phonon interactions, we perform conventional AIMD simulations in which the initial atomic structure is distorted along the eigenvector direction of IR$_x$ mode with the value of $\bf{Q}$$_x(t=0) = 1.2$\AA$\sqrt{\mu}$. 
As shown in Fig.~2(a), induced oscillations of IR$_y$ and IR$_z$ modes can be observed, indicating the anharmonic phonon couplings between the IR$_{x}$ and a linearly combined oscillation of IR$_y$ and IR$_z$.
Notably, the nonlinear phonon interaction shifts the origin of the oscillation, corresponding to a lattice distortion along $\hat{\bf{Q}}$$_{ind}=0.81\hat{\bf{Q}}$$_{y}+0.58\hat{\bf{Q}}$$_{z}$ where $\hat{\bf{Q}}$$_{y,z}$ denotes the unit vector along the direction of equilibrium IR$_{y,z}$ eigenmodes. 
Their nonlinear dynamics can also be revealed from the calculated potential energy surface as shown in Fig.~2(b). 
At equilibrium ($\bf{Q}$$_x=0$), the potential surface shows a parabolic curve with a minimum energy point at $\bf{Q}$$_{ind}^{min}=0$.
For both positive and negative phonon displacements of $\bf{Q}$$_x=\pm1.2$~\AA$\sqrt{\mu}$, the minimum points shift to a negative value ($\bf{Q}$$_{ind}^{min}<0$) regardless of the sign of $\bf{Q}$$_x$.
These results indicate that a resonant pumping of the IR$_x$ mode could strongly modify the potential surface.
Similar nonlinear phonon interactions are obtained in HgSe and $\beta$-HgS under resonant THz pumping (see Supplementary Materials (SM)~\cite{SM}).

Anharmonicity and dissipation effects of phonons should be considered to describe the non-equilibrium structural dynamics induced by a THz pulse.
To achieve this goal, we proceed with extended AIMD simulations that include direct interaction between oscillating electric field and charged ions under the Born-Oppenheimer approximation and in the periodic boundary condition with a $3 \times 3 \times 3 $ supercell geometry of HgTe (see SM for computational details, including discussion on the role of screening effect)~\cite{SM}.
In practice, the THz field excites the IR$_x$ phonon mode via an electric field oscillating: $\bf{E}$$(t)=E_0\sin(\omega (t-t_0)) e^{-(t-t_0)^2/\sigma^2}\bf{\hat{x}}$~in which $\sigma=0.8~ps$, $t_0=2.5~ps$ and $\omega=3.5$~THz with a maximum amplitude of $1.25$~MV/cm as shown in Fig.~2(c).
Under this resonant laser pumping, the oscillation amplitude of IR$_x$ is increased up to $1.8$~\AA$\sqrt{\mu}$ until $4$~ps.
Such excitation leads to the emergence of the non-zero amplitude of induced mode ($\bf{Q}$$_{ind}$) through the nonlinear phononics, as shown in Fig.~2(d).
Notably, the origin of $\bf{Q}$$_{ind}(t)$ is shifted away from the equilibrium position in the steady state with a lifetime of a few picoseconds, which is much longer than the broadening of the THz pumping pulse ($\sigma=0.8$~ps). 
Therefore, a meta-stable structural distortion is achieved through nonlinear phonon interaction between the highly excited IR$_x$ and IR$_{ind}$ modes. 
After a few picoseconds (\textgreater 8~ps), excited IR$_x$ and IR$_{ind}$ phonons dissipate via anharmonic couplings with other phonon modes, acting as a heating source and increasing the total temperature of this system. 
In the SM, we provide detailed discussions on the role of dissipation and thermostat on the light-induced meta-stable atomic geometry~\cite{SM}.

The time-averaged lattice distortion $\bf{Q}$$_{ind}^{avg}$ depends on the intensity of the applied THz pulse as shown in Fig.~2(e). 
For instance, the stronger THz pulse ($\sim1.75$~MV/cm) leads to a more considerable lattice distortion ($\bf{Q}$$^{avg}_{ind}\sim 0.84$~\AA$\sqrt{\mu}$) in HgTe, compared with the case with weaker pumping strength ($\bf{Q}$$^{avg}_{ind}\sim 0.08$~\AA$\sqrt{\mu}$ at E$=0.5$~MV/cm).
The same tendency is found in HgSe and $\beta$-HgS, which have similar nonlinear phonon interactions with that in HgTe.
These results show that the amplitude of lattice distortion in the meta-stable state can be controlled by varying the intensity of the THz field pulse in the HgTe family.

To better understand the role of anharmonicity and dissipation, we compare our extended AIMD simulation with the results from the simplified lattice model fitted to $ab~initio$ calculations~\cite{Kozina2019,Shin2021arxiv}.
This model considers harmonic IR$_x$ and IR$_{ind}$ modes and their minimal nonlinear coupling and the equations of motion are given as follows: $\ddot{Q}_x +\gamma \dot{Q}_x+ \Omega^2 Q_x = -2k_{nl} Q_x Q_{ind} + Z^*E(t)$, and $\ddot{Q}_{ind} +\gamma \dot{Q}_{ind}+ \Omega^2 Q_{ind} = -k_{nl} Q_x^2$, where $Z^*$ is the mode effective charges, $k_{nl}$ is the coupling coefficient for nonlinear phonon interaction, $\gamma$ is the dissipation rate, and $\Omega$ is the frequency of these phonon modes (see SM for details)~\cite{SM}.
We find that this model can reproduce the $ab~initio$ results about the structural evolution of HgTe very well in the low range of field strength ($E$~\textless~$1$~MV/cm) (see Fig. 2(e)). 
However, it fails to reproduce the displacements in the high field strength range ($E$~\textgreater~$1$~MV/cm), due to the restrictive description of dissipation channels and anharmonicity.
These results demonstrate that our extended AIMD calculations can provide more realistic dynamics in HgTe to investigate light-induced metastable phases than the model simulations.

\begin{figure}[t]
  \centering
  {\includegraphics[width=0.45\textwidth]{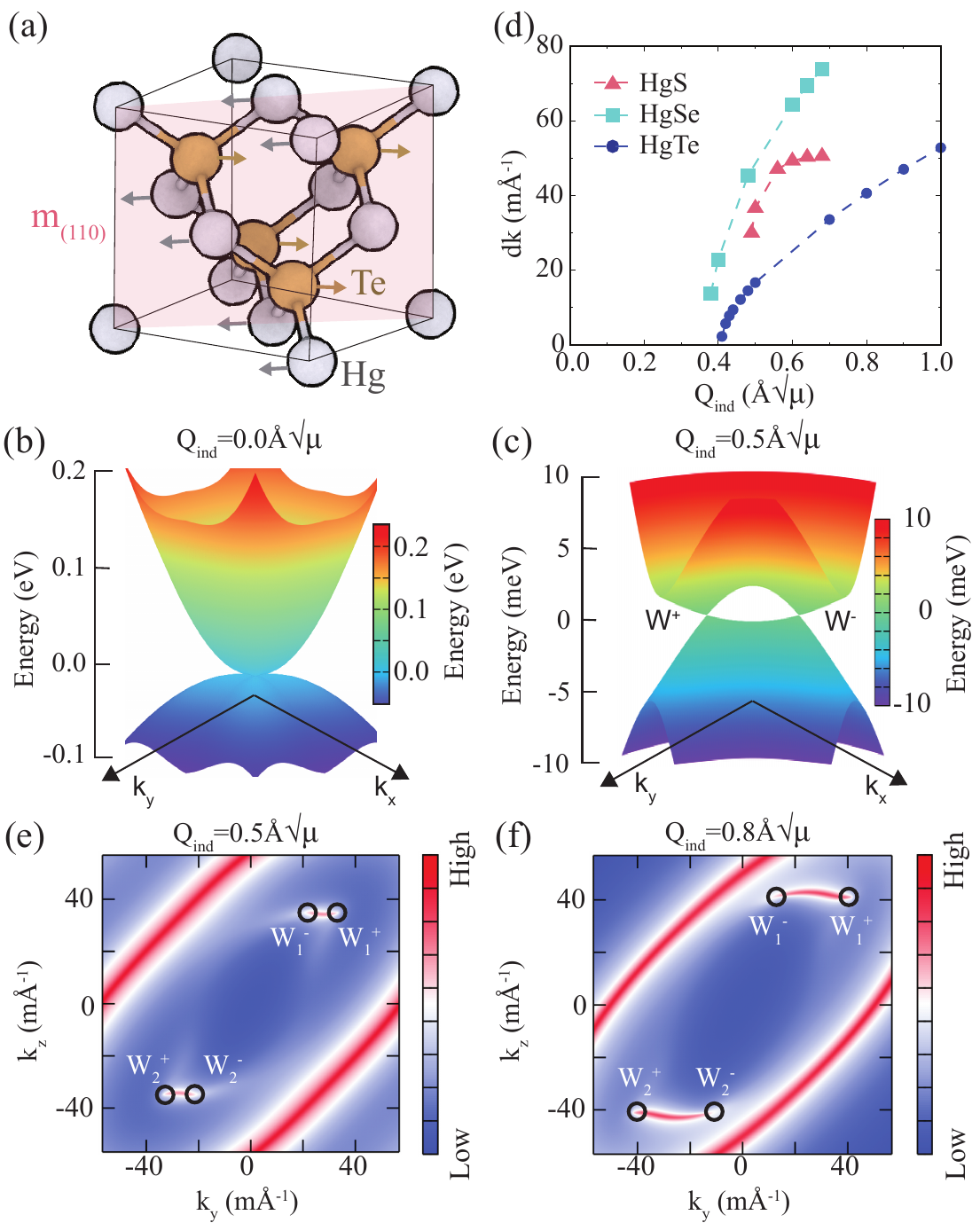}}
   \caption{
    Formation of Weyl points in the shifted equilibrium position by the nonlinear phonon interaction and possible experimental observation in TrARPES.
    (a) Displacement direction of the induced mode ($Q_{ind}$) in cubic geometry.
    (b) The band structure of HgTe with the equilibrium geometry on $k_z=0$ plane.
    (c) The band structure of HgTe in the THz-induced non-equilibrium geometry on the $k_z=35.3$~m\AA$^{-1}$ plane.
    (d) Distance between Weyl points depending on the displacement of the induced mode $Q_{ind}$ in HgTe family.
    Fermi arc surface states in the (100) surface with (e) $Q_{ind}=0.5$~\AA$\sqrt{\mu}$~and (f) $Q_{ind}=0.8$~\AA$\sqrt{\mu}$.
    In (a), pink plane indicates the (110) mirror plane.
    In (b) and (c), the color map indicates the energy variation corresponding to its z-axis.
}\label{Fig3}
\end{figure}

The THz pulse-induced structural distortion drives bulk HgTe into an ideal Weyl semimetal with four Weyl points at the Fermi level in non-equilibrium. 
At the equilibrium geometry, bulk HgTe is a gapless trivial semimetal with inverted bands, in which the degeneracy at the $\Gamma$ point (see Fig. 3(b)) is protected by the crystal symmetry ($F\bar{4}3m$ space group).
Under atomic distortion along $\bf{Q}$$_{ind}$, only one mirror symmetry $m_{1,1,0}$ (see Fig. 3(a)) is preserved in new structure ($Cm (8)$ space group). 
With a displacement of $\bf{Q}$$_{ind}=0.5$~\AA$\sqrt{\mu}$, fourfold degenerated points at the Fermi level split into four Weyl points away from the $\Gamma$ point, their positions in the momentum space are labeled as $W_1^{\pm}$ and $W_2^{\pm}$ as shown in Fig. 3(c), in which $\pm$ stands for the sign of the topological charges. 
Under mirror symmetry and time-reversal symmetry, these Weyl points can be converted from one to others (see SM~\cite{SM}), sharing the same energy eigenvalue. 
Such a phase is named the ideal Weyl semimetal~\cite{Ruan2016} since Weyl points are precisely at the Fermi level and low-energy electronic states are always topologically non-trivial. In Fig. 3(e), we plot the (100) surface state, where four Weyl points project to different positions on the side surface and Fermi arc surface states connect each pair with opposite chiral charges, which can be detected in TrARPES~\cite{Xu2015}. 
Our result shows that the THz-driven lattice distortion can induce a topological phase transition in bulk HgTe.

The positions of the light-induced Weyl points depend on the strength of the lattice distortion.
In Fig 3(d), we show that the change of distance ($d_k$) between two Weyl points $W_{1(2)}^+$ and $W_{1(2)}^-$ with respect to lattice distortions. 
Before the critical value of $\bf{Q}$$_{ind}\sim 0.4$~\AA$\sqrt{\mu}$, the nodal lines instead of Weyl points (see Fig. 3(d)) can be observed in the bulk and the similar result has been reported in strained cases~\cite{Ruan2016}. 
Above this value, for example, $\bf{Q}$$_{ind}=1.0$~\AA$\sqrt{\mu}$ corresponding to $1.75$~MV/cm maximum field strength gives rise to a Weyl point separation of $d_k=52.8$~m\AA$^{-1}$.
The same conclusion is valid for bulk HgSe and $\beta$-HgS (see Fig. 3(d)).
The evolution of the light-induced Weyl points with increasing pumping strength can be captured from the evolution of the Fermi arc surface states. 
Compared with the case for $\bf{Q}$$_{ind}=0.5$~\AA$\sqrt{\mu}$ shown in Fig. 3(e), the Fermi arc surface state (see Fig. 3(f)) becomes larger under stronger THz laser pumping ($\bf{Q}$$_{ind}=0.8$~\AA$\sqrt{\mu}$ correspondingly). We anticipate that such features will be captured in TrARPES with THz laser pumping, a powerful tool to detect the Fermi arc surface state and its related dynamic evolution~\cite{Xu2015,Yang2015,Gierz2013,zhou2023pseudospin}.

\begin{figure}[t]
  \centering
  {\includegraphics[width=0.45\textwidth]{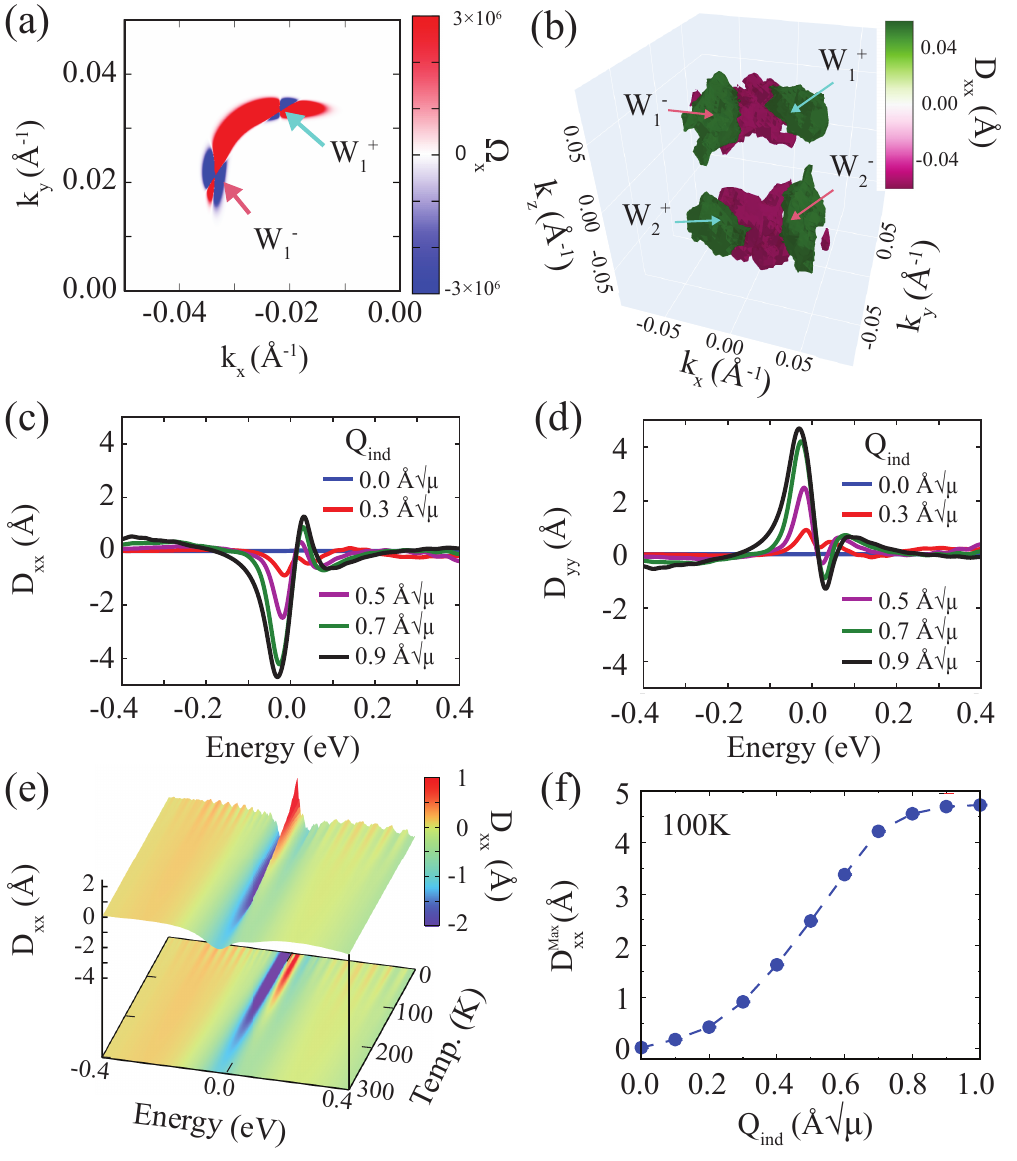}}
   \caption{
    Possible experimental observable of the light-induced Weyl points; nonlinear Hall effect.
    (a) Berry curvature $\Omega_x$ of distorted HgTe with $Q_{ind}=0.5$~\AA$\sqrt{\mu}$ at $k_z=35.3$~m\AA$^{-1}$ plane.
    (b) Berry curvature dipole of distorted HgTe with $Q_{ind}=0.5$~\AA$\sqrt{\mu}$ in 3D plot.
    Diagonal terms of Berry curvature dipole (c) $D_{xx}$ and (d) $D_{yy}$ with distorted geometry along $Q_{ind}$ at $100$~K.
    (e) Temperature and Fermi level dependency on the Berry curvature dipole $D_{xx}$ with $Q_{ind}=0.5$~\AA$\sqrt{\mu}$.
    (f) The absolute maximum value of Berry curvature dipole $D_{xx}$ near the Fermi level at $100$~K.
}\label{Fig4}
\end{figure}

Besides TrARPES, THz field-induced Weyl points can also be experimentally observed in quantum transport measurements, such as the nonlinear Hall effect~\cite{Sodemann2015,Ma2019,Zhang2018}. 
The nonlinear Hall current defined as $\bm{J}_a=\chi_{abc}\bm{E}_b\bm{E}^{*}_c$, with $\chi_{abc}=\epsilon_{adc}\frac{e^3\tau}{2(1+i\omega \tau)}D_{bd}$ depends on the Berry curvature dipole $D_{bd}$, in which $e$ is the electron charge, $\tau$ is the scattering time, and $\omega$ is the frequency of the probe laser field~\cite{Sodemann2015}.
The total Berry curvature is zero in HgTe due to the time-reversal symmetry. 
However, the inversion symmetry-breaking condition induced by the strong light-matter interaction in HgTe leads to a non-zero Berry curvature dipole that provides the nonlinear Hall effect (see Fig.~4)~\cite{Gao2020}.

The distorted geometry of HgTe induced by nonlinear phonon interaction shows a non-zero Berry curvature dipole. Based on the symmetry analysis, the bulk HgTe in equilibrium shows zero Berry curvature dipole due to inversion symmetry. 
In contrast, the distorted HgTe has a finite value with elements $D_{xx}=-D_{yy}$ and $D_{zz}=0$ constrained by the mirror symmetry (see SM~\cite{SM}). 
Furthermore, we use the density functional theory to calculate Berry curvature and Berry curvature dipole for distorted HgTe. 
In Fig. 4(a), we plot Berry curvature $\Omega$ at $k_z=35.3$~m\AA$^{-1}$ plane of Brillouin zone, which asymmetrically distributes around Weyl points and could result in a non-zero Berry curvature dipole distribution $\bf{D}$$(\textbf{k})$~\cite{Zhang2018} (see Fig. 4(b)) because $\bf{D}$$(\textbf{k})$ $\sim f_{\textbf{k}}\nabla_{\textbf{k}} \Omega(\textbf{k})$, where $f_{\textbf{k}}$ is the occupation of Bloch state. In Fig. 4(c) and 4(d), we show the calculated Berry curvature dipole for structures with different $\bf{Q}$$_{ind}$ values, corresponding to different THz pumping intensities.
Furthermore, we find that $D_{xx}$ and $D_{yy}$ have two peaks with opposite signs around the Fermi level, corresponding to contributions from electron and hole parts of Weyl fermions. 

We also investigated thermal effects on the Berry curvature dipole to verify whether the dynamic nonlinear Hall effect can be observed at finite temperatures. 
The temperature-dependent Berry curvature dipole is evaluated with an electronic Fermi-Dirac distribution function at a given temperature $T$. 
The results are shown in Fig. 4(e), whose temperature dependency indicates a measurable nonlinear Hall current can be experimentally achieved at high-temperature regions (even at room temperature). 
Once we fix the temperature (see Fig. 4(f)), peak values of Berry curvature dipole increase with the enhancement of pumping strength (larger $\bf{Q}$$_{ind}$ value) because of the increasing distance between Weyl points with opposite chiral charges. 
These results indicate the possibility of identifying light-induced Weyl points in bulk HgTe via pump-probe ultrafast experiments.

In summary, we theoretically investigate a light-induced topological phase transition in the family of HgTe materials through nonlinear phonon interaction.
We use an extended AIMD framework to show that THz laser coherent pumping could drive a new non-equilibrium steady-state with a distorted atomic structure via nonlinear phononics. 
The long-lived lattice distortion is enhanced by increasing the pump laser strength. 
Bulk HgTe with the distorted geometry hosts four Weyl points at the Fermi level, an ideal Weyl semimetal in non-equilibrium. 
The same scenario has been observed in HgSe and $\beta$-HgS. 
We anticipate that ultrafast pump-probe experiments can be used to validate the proposed topological phase transitions in bulk HgTe family via detecting surface Fermi arc states and nonlinear Hall effects. 
Our work provides an opportunity to optically manipulate topological properties of quantum materials via nonlinear phononics, which is important for future ultrafast all-optical topological device switching. 
Furthermore, the topics discussed herein can be extended to the new field of cavity materials engineering or cavity-induced phase transition~\cite{Basov2020,Bloch2022,Simone2021,Hannes2021,dante2022}. 

\begin{acknowledgments}
We acknowledge financial support from the Grupos Consolidados (IT1453-22), the Cluster of Excellence `CUI: Advanced Imaging of Matter' of the Deutsche Forschungsgemeinschaft (DFG) - EXC 2056 - project ID 390715994. 
D.S. was supported by the National Research Foundation of Korea (NRF) grant funded by the Korea government (MSIT) (No. RS-2023-00253716). The computational work was supported by the National Supercomputing Center with supercomputing resources including technical support (KSC-2023-CRE-0073). 
P.Z.T. was supported by the National Natural Science Foundation of China (Grants No. 12234011 and No. 12374053).
We also acknowledge support from the Max Planck–New York Center for Non-Equilibrium Quantum Phenomena.
The Flatiron Institute is a division of the Simons Foundation. 
The authors thank Dante M. Kennes and Noejung Park for the fruitful discussion.
\end{acknowledgments}


\begin{thebibliography}{56}%
\makeatletter
\providecommand \@ifxundefined [1]{%
 \@ifx{#1\undefined}
}%
\providecommand \@ifnum [1]{%
 \ifnum #1\expandafter \@firstoftwo
 \else \expandafter \@secondoftwo
 \fi
}%
\providecommand \@ifx [1]{%
 \ifx #1\expandafter \@firstoftwo
 \else \expandafter \@secondoftwo
 \fi
}%
\providecommand \natexlab [1]{#1}%
\providecommand \enquote  [1]{``#1''}%
\providecommand \bibnamefont  [1]{#1}%
\providecommand \bibfnamefont [1]{#1}%
\providecommand \citenamefont [1]{#1}%
\providecommand \href@noop [0]{\@secondoftwo}%
\providecommand \href [0]{\begingroup \@sanitize@url \@href}%
\providecommand \@href[1]{\@@startlink{#1}\@@href}%
\providecommand \@@href[1]{\endgroup#1\@@endlink}%
\providecommand \@sanitize@url [0]{\catcode `\\12\catcode `\$12\catcode
  `\&12\catcode `\#12\catcode `\^12\catcode `\_12\catcode `\%12\relax}%
\providecommand \@@startlink[1]{}%
\providecommand \@@endlink[0]{}%
\providecommand \url  [0]{\begingroup\@sanitize@url \@url }%
\providecommand \@url [1]{\endgroup\@href {#1}{\urlprefix }}%
\providecommand \urlprefix  [0]{URL }%
\providecommand \Eprint [0]{\href }%
\providecommand \doibase [0]{https://doi.org/}%
\providecommand \selectlanguage [0]{\@gobble}%
\providecommand \bibinfo  [0]{\@secondoftwo}%
\providecommand \bibfield  [0]{\@secondoftwo}%
\providecommand \translation [1]{[#1]}%
\providecommand \BibitemOpen [0]{}%
\providecommand \bibitemStop [0]{}%
\providecommand \bibitemNoStop [0]{.\EOS\space}%
\providecommand \EOS [0]{\spacefactor3000\relax}%
\providecommand \BibitemShut  [1]{\csname bibitem#1\endcsname}%
\let\auto@bib@innerbib\@empty
\bibitem [{\citenamefont {de~la Torre}\ \emph {et~al.}(2021)\citenamefont
  {de~la Torre}, \citenamefont {Kennes}, \citenamefont {Claassen},
  \citenamefont {Gerber}, \citenamefont {McIver},\ and\ \citenamefont
  {Sentef}}]{Terro2021}%
  \BibitemOpen
  \bibfield  {author} {\bibinfo {author} {\bibfnamefont {A.}~\bibnamefont
  {de~la Torre}}, \bibinfo {author} {\bibfnamefont {D.~M.}\ \bibnamefont
  {Kennes}}, \bibinfo {author} {\bibfnamefont {M.}~\bibnamefont {Claassen}},
  \bibinfo {author} {\bibfnamefont {S.}~\bibnamefont {Gerber}}, \bibinfo
  {author} {\bibfnamefont {J.~W.}\ \bibnamefont {McIver}},\ and\ \bibinfo
  {author} {\bibfnamefont {M.~A.}\ \bibnamefont {Sentef}},\ }\href@noop {}
  {\bibfield  {journal} {\bibinfo  {journal} {Rev. Mod. Phys.}\ }\textbf
  {\bibinfo {volume} {93}},\ \bibinfo {pages} {041002} (\bibinfo {year}
  {2021})}\BibitemShut {NoStop}%
\bibitem [{\citenamefont {Kennes}\ and\ \citenamefont
  {Rubio}(2022)}]{dante2022}%
  \BibitemOpen
  \bibfield  {author} {\bibinfo {author} {\bibfnamefont {D.~M.}\ \bibnamefont
  {Kennes}}\ and\ \bibinfo {author} {\bibfnamefont {A.}~\bibnamefont {Rubio}},\
  }\href@noop {} {\bibfield  {journal} {\bibinfo  {journal} {arXiv}\ }
  (\bibinfo {year} {2022})},\ \Eprint {https://arxiv.org/abs/2204.11928}
  {2204.11928} \BibitemShut {NoStop}%
\bibitem [{\citenamefont {Bao}\ \emph {et~al.}(2022)\citenamefont {Bao},
  \citenamefont {Tang}, \citenamefont {Sun},\ and\ \citenamefont
  {Zhou}}]{Bao2022}%
  \BibitemOpen
  \bibfield  {author} {\bibinfo {author} {\bibfnamefont {C.}~\bibnamefont
  {Bao}}, \bibinfo {author} {\bibfnamefont {P.}~\bibnamefont {Tang}}, \bibinfo
  {author} {\bibfnamefont {D.}~\bibnamefont {Sun}},\ and\ \bibinfo {author}
  {\bibfnamefont {S.}~\bibnamefont {Zhou}},\ }\href@noop {} {\bibfield
  {journal} {\bibinfo  {journal} {Nat. Rev. Phys.}\ }\textbf {\bibinfo {volume}
  {4}},\ \bibinfo {pages} {33} (\bibinfo {year} {2022})}\BibitemShut {NoStop}%
\bibitem [{\citenamefont {Disa}\ \emph {et~al.}(2021)\citenamefont {Disa},
  \citenamefont {Nova},\ and\ \citenamefont {Cavalleri}}]{disa2021engineering}%
  \BibitemOpen
  \bibfield  {author} {\bibinfo {author} {\bibfnamefont {A.~S.}\ \bibnamefont
  {Disa}}, \bibinfo {author} {\bibfnamefont {T.~F.}\ \bibnamefont {Nova}},\
  and\ \bibinfo {author} {\bibfnamefont {A.}~\bibnamefont {Cavalleri}},\
  }\href@noop {} {\bibfield  {journal} {\bibinfo  {journal} {Nat. Phys.}\
  }\textbf {\bibinfo {volume} {17}},\ \bibinfo {pages} {1087} (\bibinfo {year}
  {2021})}\BibitemShut {NoStop}%
\bibitem [{\citenamefont {Basov}\ \emph {et~al.}(2021)\citenamefont {Basov},
  \citenamefont {Asenjo-Garcia}, \citenamefont {Schuck}, \citenamefont {Zhu},\
  and\ \citenamefont {Rubio}}]{Basov2020}%
  \BibitemOpen
  \bibfield  {author} {\bibinfo {author} {\bibfnamefont {D.~N.}\ \bibnamefont
  {Basov}}, \bibinfo {author} {\bibfnamefont {A.}~\bibnamefont
  {Asenjo-Garcia}}, \bibinfo {author} {\bibfnamefont {P.~J.}\ \bibnamefont
  {Schuck}}, \bibinfo {author} {\bibfnamefont {X.}~\bibnamefont {Zhu}},\ and\
  \bibinfo {author} {\bibfnamefont {A.}~\bibnamefont {Rubio}},\ }\href@noop {}
  {\bibfield  {journal} {\bibinfo  {journal} {Nanophotonics}\ }\textbf
  {\bibinfo {volume} {10}},\ \bibinfo {pages} {549} (\bibinfo {year}
  {2021})}\BibitemShut {NoStop}%
\bibitem [{\citenamefont {Bloch}\ \emph {et~al.}(2022)\citenamefont {Bloch},
  \citenamefont {Cavalleri}, \citenamefont {Galitski}, \citenamefont {Hafezi},\
  and\ \citenamefont {Rubio}}]{Bloch2022}%
  \BibitemOpen
  \bibfield  {author} {\bibinfo {author} {\bibfnamefont {J.}~\bibnamefont
  {Bloch}}, \bibinfo {author} {\bibfnamefont {A.}~\bibnamefont {Cavalleri}},
  \bibinfo {author} {\bibfnamefont {V.}~\bibnamefont {Galitski}}, \bibinfo
  {author} {\bibfnamefont {M.}~\bibnamefont {Hafezi}},\ and\ \bibinfo {author}
  {\bibfnamefont {A.}~\bibnamefont {Rubio}},\ }\href
  {https://doi.org/10.1038/s41586-022-04726-w} {\bibfield  {journal} {\bibinfo
  {journal} {Nature}\ }\textbf {\bibinfo {volume} {606}},\ \bibinfo {pages}
  {41} (\bibinfo {year} {2022})}\BibitemShut {NoStop}%
\bibitem [{\citenamefont {Oka}\ and\ \citenamefont {Kitamura}(2018)}]{oka2018}%
  \BibitemOpen
  \bibfield  {author} {\bibinfo {author} {\bibfnamefont {T.}~\bibnamefont
  {Oka}}\ and\ \bibinfo {author} {\bibfnamefont {S.}~\bibnamefont {Kitamura}},\
  }\href@noop {} {\bibfield  {journal} {\bibinfo  {journal} {Annu. Rev.
  Condens. Matter Phys.}\ }\textbf {\bibinfo {volume} {10}},\ \bibinfo {pages}
  {1} (\bibinfo {year} {2018})}\BibitemShut {NoStop}%
\bibitem [{\citenamefont {Hubener}\ \emph {et~al.}(2021)\citenamefont
  {Hubener}, \citenamefont {Giovannini}, \citenamefont {Schäfer},
  \citenamefont {Andberger}, \citenamefont {Ruggenthaler}, \citenamefont
  {Faist},\ and\ \citenamefont {Rubio}}]{Hannes2021}%
  \BibitemOpen
  \bibfield  {author} {\bibinfo {author} {\bibfnamefont {H.}~\bibnamefont
  {Hubener}}, \bibinfo {author} {\bibfnamefont {U.~D.}\ \bibnamefont
  {Giovannini}}, \bibinfo {author} {\bibfnamefont {C.}~\bibnamefont
  {Schäfer}}, \bibinfo {author} {\bibfnamefont {J.}~\bibnamefont {Andberger}},
  \bibinfo {author} {\bibfnamefont {M.}~\bibnamefont {Ruggenthaler}}, \bibinfo
  {author} {\bibfnamefont {J.}~\bibnamefont {Faist}},\ and\ \bibinfo {author}
  {\bibfnamefont {A.}~\bibnamefont {Rubio}},\ }\href@noop {} {\bibfield
  {journal} {\bibinfo  {journal} {Nat. Mater.}\ }\textbf {\bibinfo {volume}
  {20}},\ \bibinfo {pages} {438} (\bibinfo {year} {2021})}\BibitemShut
  {NoStop}%
\bibitem [{\citenamefont {Ruggenthaler}\ \emph {et~al.}(2018)\citenamefont
  {Ruggenthaler}, \citenamefont {Tancogne-Dejean}, \citenamefont {Flick},
  \citenamefont {Appel},\ and\ \citenamefont {Rubio}}]{Rugg2018}%
  \BibitemOpen
  \bibfield  {author} {\bibinfo {author} {\bibfnamefont {M.}~\bibnamefont
  {Ruggenthaler}}, \bibinfo {author} {\bibfnamefont {N.}~\bibnamefont
  {Tancogne-Dejean}}, \bibinfo {author} {\bibfnamefont {J.}~\bibnamefont
  {Flick}}, \bibinfo {author} {\bibfnamefont {H.}~\bibnamefont {Appel}},\ and\
  \bibinfo {author} {\bibfnamefont {A.}~\bibnamefont {Rubio}},\ }\href@noop {}
  {\bibfield  {journal} {\bibinfo  {journal} {Nat. Rev. Chem.}\ }\textbf
  {\bibinfo {volume} {2}},\ \bibinfo {pages} {0118} (\bibinfo {year}
  {2018})}\BibitemShut {NoStop}%
\bibitem [{\citenamefont {Zhou}\ \emph
  {et~al.}(2023{\natexlab{a}})\citenamefont {Zhou}, \citenamefont {Bao},
  \citenamefont {Fan}, \citenamefont {Zhou}, \citenamefont {Gao}, \citenamefont
  {Zhong}, \citenamefont {Lin}, \citenamefont {Liu}, \citenamefont {Yu},
  \citenamefont {Tang}, \citenamefont {Meng}, \citenamefont {Duan},\ and\
  \citenamefont {Zhou}}]{zhou2023pseudospin}%
  \BibitemOpen
  \bibfield  {author} {\bibinfo {author} {\bibfnamefont {S.}~\bibnamefont
  {Zhou}}, \bibinfo {author} {\bibfnamefont {C.}~\bibnamefont {Bao}}, \bibinfo
  {author} {\bibfnamefont {B.}~\bibnamefont {Fan}}, \bibinfo {author}
  {\bibfnamefont {H.}~\bibnamefont {Zhou}}, \bibinfo {author} {\bibfnamefont
  {Q.}~\bibnamefont {Gao}}, \bibinfo {author} {\bibfnamefont {H.}~\bibnamefont
  {Zhong}}, \bibinfo {author} {\bibfnamefont {T.}~\bibnamefont {Lin}}, \bibinfo
  {author} {\bibfnamefont {H.}~\bibnamefont {Liu}}, \bibinfo {author}
  {\bibfnamefont {P.}~\bibnamefont {Yu}}, \bibinfo {author} {\bibfnamefont
  {P.}~\bibnamefont {Tang}}, \bibinfo {author} {\bibfnamefont {S.}~\bibnamefont
  {Meng}}, \bibinfo {author} {\bibfnamefont {W.}~\bibnamefont {Duan}},\ and\
  \bibinfo {author} {\bibfnamefont {S.}~\bibnamefont {Zhou}},\ }\href@noop {}
  {\bibfield  {journal} {\bibinfo  {journal} {Nature}\ }\textbf {\bibinfo
  {volume} {614}},\ \bibinfo {pages} {75} (\bibinfo {year}
  {2023}{\natexlab{a}})}\BibitemShut {NoStop}%
\bibitem [{\citenamefont {Zhou}\ \emph
  {et~al.}(2023{\natexlab{b}})\citenamefont {Zhou}, \citenamefont {Bao},
  \citenamefont {Fan}, \citenamefont {Wang}, \citenamefont {Zhong},
  \citenamefont {Zhang}, \citenamefont {Tang}, \citenamefont {Duan},\ and\
  \citenamefont {Zhou}}]{ZhouBP2023_PRL}%
  \BibitemOpen
  \bibfield  {author} {\bibinfo {author} {\bibfnamefont {S.}~\bibnamefont
  {Zhou}}, \bibinfo {author} {\bibfnamefont {C.}~\bibnamefont {Bao}}, \bibinfo
  {author} {\bibfnamefont {B.}~\bibnamefont {Fan}}, \bibinfo {author}
  {\bibfnamefont {F.}~\bibnamefont {Wang}}, \bibinfo {author} {\bibfnamefont
  {H.}~\bibnamefont {Zhong}}, \bibinfo {author} {\bibfnamefont
  {H.}~\bibnamefont {Zhang}}, \bibinfo {author} {\bibfnamefont
  {P.}~\bibnamefont {Tang}}, \bibinfo {author} {\bibfnamefont {W.}~\bibnamefont
  {Duan}},\ and\ \bibinfo {author} {\bibfnamefont {S.}~\bibnamefont {Zhou}},\
  }\href {https://doi.org/10.1103/PhysRevLett.131.116401} {\bibfield  {journal}
  {\bibinfo  {journal} {Phys. Rev. Lett.}\ }\textbf {\bibinfo {volume} {131}},\
  \bibinfo {pages} {116401} (\bibinfo {year} {2023}{\natexlab{b}})}\BibitemShut
  {NoStop}%
\bibitem [{\citenamefont {Rudner}\ and\ \citenamefont
  {Lindner}(2020)}]{rudner2020band}%
  \BibitemOpen
  \bibfield  {author} {\bibinfo {author} {\bibfnamefont {M.~S.}\ \bibnamefont
  {Rudner}}\ and\ \bibinfo {author} {\bibfnamefont {N.~H.}\ \bibnamefont
  {Lindner}},\ }\href@noop {} {\bibfield  {journal} {\bibinfo  {journal} {Nat.
  Rev. Phys.}\ }\textbf {\bibinfo {volume} {2}},\ \bibinfo {pages} {229}
  (\bibinfo {year} {2020})}\BibitemShut {NoStop}%
\bibitem [{\citenamefont {Wang}\ \emph {et~al.}(2013)\citenamefont {Wang},
  \citenamefont {Steinberg}, \citenamefont {Jarillo-Herrero},\ and\
  \citenamefont {Gedik}}]{Wang2013}%
  \BibitemOpen
  \bibfield  {author} {\bibinfo {author} {\bibfnamefont {Y.~H.}\ \bibnamefont
  {Wang}}, \bibinfo {author} {\bibfnamefont {H.}~\bibnamefont {Steinberg}},
  \bibinfo {author} {\bibfnamefont {P.}~\bibnamefont {Jarillo-Herrero}},\ and\
  \bibinfo {author} {\bibfnamefont {N.}~\bibnamefont {Gedik}},\ }\href@noop {}
  {\bibfield  {journal} {\bibinfo  {journal} {Science}\ }\textbf {\bibinfo
  {volume} {342}},\ \bibinfo {pages} {453} (\bibinfo {year}
  {2013})}\BibitemShut {NoStop}%
\bibitem [{\citenamefont {Aeschlimann}\ \emph {et~al.}(2021)\citenamefont
  {Aeschlimann}, \citenamefont {Sato}, \citenamefont {Krause}, \citenamefont
  {Ch{\'a}vez-Cervantes}, \citenamefont {De~Giovannini}, \citenamefont
  {H{\"u}bener}, \citenamefont {Forti}, \citenamefont {Coletti}, \citenamefont
  {Hanff}, \citenamefont {Rossnagel}, \citenamefont {Rubio},\ and\
  \citenamefont {Gierz}}]{GierzWSe2}%
  \BibitemOpen
  \bibfield  {author} {\bibinfo {author} {\bibfnamefont {S.}~\bibnamefont
  {Aeschlimann}}, \bibinfo {author} {\bibfnamefont {S.~A.}\ \bibnamefont
  {Sato}}, \bibinfo {author} {\bibfnamefont {R.}~\bibnamefont {Krause}},
  \bibinfo {author} {\bibfnamefont {M.}~\bibnamefont {Ch{\'a}vez-Cervantes}},
  \bibinfo {author} {\bibfnamefont {U.}~\bibnamefont {De~Giovannini}}, \bibinfo
  {author} {\bibfnamefont {H.}~\bibnamefont {H{\"u}bener}}, \bibinfo {author}
  {\bibfnamefont {S.}~\bibnamefont {Forti}}, \bibinfo {author} {\bibfnamefont
  {C.}~\bibnamefont {Coletti}}, \bibinfo {author} {\bibfnamefont
  {K.}~\bibnamefont {Hanff}}, \bibinfo {author} {\bibfnamefont
  {K.}~\bibnamefont {Rossnagel}}, \bibinfo {author} {\bibfnamefont
  {A.}~\bibnamefont {Rubio}},\ and\ \bibinfo {author} {\bibfnamefont
  {I.}~\bibnamefont {Gierz}},\ }\href@noop {} {\bibfield  {journal} {\bibinfo
  {journal} {Nano Lett.}\ }\textbf {\bibinfo {volume} {21}},\ \bibinfo {pages}
  {5028} (\bibinfo {year} {2021})}\BibitemShut {NoStop}%
\bibitem [{\citenamefont {Iadecola}\ \emph {et~al.}(2015)\citenamefont
  {Iadecola}, \citenamefont {Neupert},\ and\ \citenamefont
  {Chamon}}]{Iadecola2015}%
  \BibitemOpen
  \bibfield  {author} {\bibinfo {author} {\bibfnamefont {T.}~\bibnamefont
  {Iadecola}}, \bibinfo {author} {\bibfnamefont {T.}~\bibnamefont {Neupert}},\
  and\ \bibinfo {author} {\bibfnamefont {C.}~\bibnamefont {Chamon}},\
  }\href@noop {} {\bibfield  {journal} {\bibinfo  {journal} {Phys. Rev. B}\
  }\textbf {\bibinfo {volume} {91}},\ \bibinfo {pages} {235133} (\bibinfo
  {year} {2015})}\BibitemShut {NoStop}%
\bibitem [{\citenamefont {Shin}\ \emph {et~al.}(2020)\citenamefont {Shin},
  \citenamefont {Sato}, \citenamefont {Hübener}, \citenamefont {Giovannini},
  \citenamefont {Park},\ and\ \citenamefont {Rubio}}]{Shin2020}%
  \BibitemOpen
  \bibfield  {author} {\bibinfo {author} {\bibfnamefont {D.}~\bibnamefont
  {Shin}}, \bibinfo {author} {\bibfnamefont {S.~A.}\ \bibnamefont {Sato}},
  \bibinfo {author} {\bibfnamefont {H.}~\bibnamefont {Hübener}}, \bibinfo
  {author} {\bibfnamefont {U.~D.}\ \bibnamefont {Giovannini}}, \bibinfo
  {author} {\bibfnamefont {N.}~\bibnamefont {Park}},\ and\ \bibinfo {author}
  {\bibfnamefont {A.}~\bibnamefont {Rubio}},\ }\href@noop {} {\bibfield
  {journal} {\bibinfo  {journal} {Npj Comput. Mater.}\ }\textbf {\bibinfo
  {volume} {6}},\ \bibinfo {pages} {182} (\bibinfo {year} {2020})}\BibitemShut
  {NoStop}%
\bibitem [{\citenamefont {F{\"o}rst}\ \emph {et~al.}(2011)\citenamefont
  {F{\"o}rst}, \citenamefont {Manzoni}, \citenamefont {Kaiser}, \citenamefont
  {Tomioka}, \citenamefont {Tokura}, \citenamefont {Merlin},\ and\
  \citenamefont {Cavalleri}}]{forst2011nonlinear}%
  \BibitemOpen
  \bibfield  {author} {\bibinfo {author} {\bibfnamefont {M.}~\bibnamefont
  {F{\"o}rst}}, \bibinfo {author} {\bibfnamefont {C.}~\bibnamefont {Manzoni}},
  \bibinfo {author} {\bibfnamefont {S.}~\bibnamefont {Kaiser}}, \bibinfo
  {author} {\bibfnamefont {Y.}~\bibnamefont {Tomioka}}, \bibinfo {author}
  {\bibfnamefont {Y.-n.}\ \bibnamefont {Tokura}}, \bibinfo {author}
  {\bibfnamefont {R.}~\bibnamefont {Merlin}},\ and\ \bibinfo {author}
  {\bibfnamefont {A.}~\bibnamefont {Cavalleri}},\ }\href@noop {} {\bibfield
  {journal} {\bibinfo  {journal} {Nat. Phys.}\ }\textbf {\bibinfo {volume}
  {7}},\ \bibinfo {pages} {854} (\bibinfo {year} {2011})}\BibitemShut {NoStop}%
\bibitem [{\citenamefont {Mitrano}\ \emph {et~al.}(2016)\citenamefont
  {Mitrano}, \citenamefont {Cantaluppi}, \citenamefont {Nicoletti},
  \citenamefont {Kaiser}, \citenamefont {Perucchi}, \citenamefont {Lupi},
  \citenamefont {Pietro}, \citenamefont {Pontiroli}, \citenamefont {Riccò},
  \citenamefont {Subedi}, \citenamefont {Clark}, \citenamefont {Jaksch},\ and\
  \citenamefont {Cavalleri}}]{Mitrano2016}%
  \BibitemOpen
  \bibfield  {author} {\bibinfo {author} {\bibfnamefont {M.}~\bibnamefont
  {Mitrano}}, \bibinfo {author} {\bibfnamefont {A.}~\bibnamefont {Cantaluppi}},
  \bibinfo {author} {\bibfnamefont {D.}~\bibnamefont {Nicoletti}}, \bibinfo
  {author} {\bibfnamefont {S.}~\bibnamefont {Kaiser}}, \bibinfo {author}
  {\bibfnamefont {A.}~\bibnamefont {Perucchi}}, \bibinfo {author}
  {\bibfnamefont {S.}~\bibnamefont {Lupi}}, \bibinfo {author} {\bibfnamefont
  {P.~D.}\ \bibnamefont {Pietro}}, \bibinfo {author} {\bibfnamefont
  {D.}~\bibnamefont {Pontiroli}}, \bibinfo {author} {\bibfnamefont
  {M.}~\bibnamefont {Riccò}}, \bibinfo {author} {\bibfnamefont
  {A.}~\bibnamefont {Subedi}}, \bibinfo {author} {\bibfnamefont {S.~R.}\
  \bibnamefont {Clark}}, \bibinfo {author} {\bibfnamefont {D.}~\bibnamefont
  {Jaksch}},\ and\ \bibinfo {author} {\bibfnamefont {A.}~\bibnamefont
  {Cavalleri}},\ }\href@noop {} {\bibfield  {journal} {\bibinfo  {journal}
  {Nature}\ }\textbf {\bibinfo {volume} {530}},\ \bibinfo {pages} {461}
  (\bibinfo {year} {2016})}\BibitemShut {NoStop}%
\bibitem [{\citenamefont {Buzzi}\ \emph {et~al.}(2020)\citenamefont {Buzzi},
  \citenamefont {Nicoletti}, \citenamefont {Fechner}, \citenamefont
  {Tancogne-Dejean}, \citenamefont {Sentef}, \citenamefont {Georges},
  \citenamefont {Biesner}, \citenamefont {Uykur}, \citenamefont {Dressel},
  \citenamefont {Henderson}, \citenamefont {Siegrist}, \citenamefont
  {Schlueter}, \citenamefont {Miyagawa}, \citenamefont {Kanoda}, \citenamefont
  {Nam}, \citenamefont {Ardavan}, \citenamefont {Coulthard}, \citenamefont
  {Tindall}, \citenamefont {Schlawin}, \citenamefont {Jaksch},\ and\
  \citenamefont {Cavalleri}}]{Buzzi2020}%
  \BibitemOpen
  \bibfield  {author} {\bibinfo {author} {\bibfnamefont {M.}~\bibnamefont
  {Buzzi}}, \bibinfo {author} {\bibfnamefont {D.}~\bibnamefont {Nicoletti}},
  \bibinfo {author} {\bibfnamefont {M.}~\bibnamefont {Fechner}}, \bibinfo
  {author} {\bibfnamefont {N.}~\bibnamefont {Tancogne-Dejean}}, \bibinfo
  {author} {\bibfnamefont {M.~A.}\ \bibnamefont {Sentef}}, \bibinfo {author}
  {\bibfnamefont {A.}~\bibnamefont {Georges}}, \bibinfo {author} {\bibfnamefont
  {T.}~\bibnamefont {Biesner}}, \bibinfo {author} {\bibfnamefont
  {E.}~\bibnamefont {Uykur}}, \bibinfo {author} {\bibfnamefont
  {M.}~\bibnamefont {Dressel}}, \bibinfo {author} {\bibfnamefont
  {A.}~\bibnamefont {Henderson}}, \bibinfo {author} {\bibfnamefont
  {T.}~\bibnamefont {Siegrist}}, \bibinfo {author} {\bibfnamefont {J.~A.}\
  \bibnamefont {Schlueter}}, \bibinfo {author} {\bibfnamefont {K.}~\bibnamefont
  {Miyagawa}}, \bibinfo {author} {\bibfnamefont {K.}~\bibnamefont {Kanoda}},
  \bibinfo {author} {\bibfnamefont {M.-S.}\ \bibnamefont {Nam}}, \bibinfo
  {author} {\bibfnamefont {A.}~\bibnamefont {Ardavan}}, \bibinfo {author}
  {\bibfnamefont {J.}~\bibnamefont {Coulthard}}, \bibinfo {author}
  {\bibfnamefont {J.}~\bibnamefont {Tindall}}, \bibinfo {author} {\bibfnamefont
  {F.}~\bibnamefont {Schlawin}}, \bibinfo {author} {\bibfnamefont
  {D.}~\bibnamefont {Jaksch}},\ and\ \bibinfo {author} {\bibfnamefont
  {A.}~\bibnamefont {Cavalleri}},\ }\href@noop {} {\bibfield  {journal}
  {\bibinfo  {journal} {Phys. Rev. X}\ }\textbf {\bibinfo {volume} {10}},\
  \bibinfo {pages} {031028} (\bibinfo {year} {2020})}\BibitemShut {NoStop}%
\bibitem [{\citenamefont {Kozina}\ \emph {et~al.}(2019)\citenamefont {Kozina},
  \citenamefont {Fechner}, \citenamefont {Marsik}, \citenamefont {Driel},
  \citenamefont {Glownia}, \citenamefont {Bernhard}, \citenamefont {Radovic},
  \citenamefont {Zhu}, \citenamefont {Bonetti}, \citenamefont {Staub},\ and\
  \citenamefont {Hoffmann}}]{Kozina2019}%
  \BibitemOpen
  \bibfield  {author} {\bibinfo {author} {\bibfnamefont {M.}~\bibnamefont
  {Kozina}}, \bibinfo {author} {\bibfnamefont {M.}~\bibnamefont {Fechner}},
  \bibinfo {author} {\bibfnamefont {P.}~\bibnamefont {Marsik}}, \bibinfo
  {author} {\bibfnamefont {T.~v.}\ \bibnamefont {Driel}}, \bibinfo {author}
  {\bibfnamefont {J.~M.}\ \bibnamefont {Glownia}}, \bibinfo {author}
  {\bibfnamefont {C.}~\bibnamefont {Bernhard}}, \bibinfo {author}
  {\bibfnamefont {M.}~\bibnamefont {Radovic}}, \bibinfo {author} {\bibfnamefont
  {D.}~\bibnamefont {Zhu}}, \bibinfo {author} {\bibfnamefont {S.}~\bibnamefont
  {Bonetti}}, \bibinfo {author} {\bibfnamefont {U.}~\bibnamefont {Staub}},\
  and\ \bibinfo {author} {\bibfnamefont {M.~C.}\ \bibnamefont {Hoffmann}},\
  }\href@noop {} {\bibfield  {journal} {\bibinfo  {journal} {Nat. Phys.}\
  }\textbf {\bibinfo {volume} {15}},\ \bibinfo {pages} {387} (\bibinfo {year}
  {2019})}\BibitemShut {NoStop}%
\bibitem [{\citenamefont {Subedi}(2015)}]{Subedi2015}%
  \BibitemOpen
  \bibfield  {author} {\bibinfo {author} {\bibfnamefont {A.}~\bibnamefont
  {Subedi}},\ }\href@noop {} {\bibfield  {journal} {\bibinfo  {journal} {Phys.
  Rev. B}\ }\textbf {\bibinfo {volume} {92}},\ \bibinfo {pages} {214303}
  (\bibinfo {year} {2015})}\BibitemShut {NoStop}%
\bibitem [{\citenamefont {Mankowsky}\ \emph {et~al.}(2016)\citenamefont
  {Mankowsky}, \citenamefont {Först},\ and\ \citenamefont
  {Cavalleri}}]{Mankowsky2016}%
  \BibitemOpen
  \bibfield  {author} {\bibinfo {author} {\bibfnamefont {R.}~\bibnamefont
  {Mankowsky}}, \bibinfo {author} {\bibfnamefont {M.}~\bibnamefont {Först}},\
  and\ \bibinfo {author} {\bibfnamefont {A.}~\bibnamefont {Cavalleri}},\
  }\href@noop {} {\bibfield  {journal} {\bibinfo  {journal} {Rep. Prog. Phys.}\
  }\textbf {\bibinfo {volume} {79}},\ \bibinfo {pages} {064503} (\bibinfo
  {year} {2016})}\BibitemShut {NoStop}%
\bibitem [{\citenamefont {Basov}\ \emph {et~al.}(2017)\citenamefont {Basov},
  \citenamefont {Averitt},\ and\ \citenamefont {Hsieh}}]{Basov2017}%
  \BibitemOpen
  \bibfield  {author} {\bibinfo {author} {\bibfnamefont {D.~N.}\ \bibnamefont
  {Basov}}, \bibinfo {author} {\bibfnamefont {R.~D.}\ \bibnamefont {Averitt}},\
  and\ \bibinfo {author} {\bibfnamefont {D.}~\bibnamefont {Hsieh}},\
  }\href@noop {} {\bibfield  {journal} {\bibinfo  {journal} {Nat. Mater.}\
  }\textbf {\bibinfo {volume} {16}},\ \bibinfo {pages} {1077} (\bibinfo {year}
  {2017})}\BibitemShut {NoStop}%
\bibitem [{\citenamefont {Fausti}\ \emph {et~al.}(2011)\citenamefont {Fausti},
  \citenamefont {Tobey}, \citenamefont {Dean}, \citenamefont {Kaiser},
  \citenamefont {Dienst}, \citenamefont {Hoffmann}, \citenamefont {Pyon},
  \citenamefont {Takayama}, \citenamefont {Takagi},\ and\ \citenamefont
  {Cavalleri}}]{Fausti2011}%
  \BibitemOpen
  \bibfield  {author} {\bibinfo {author} {\bibfnamefont {D.}~\bibnamefont
  {Fausti}}, \bibinfo {author} {\bibfnamefont {R.~I.}\ \bibnamefont {Tobey}},
  \bibinfo {author} {\bibfnamefont {N.}~\bibnamefont {Dean}}, \bibinfo {author}
  {\bibfnamefont {S.}~\bibnamefont {Kaiser}}, \bibinfo {author} {\bibfnamefont
  {A.}~\bibnamefont {Dienst}}, \bibinfo {author} {\bibfnamefont {M.~C.}\
  \bibnamefont {Hoffmann}}, \bibinfo {author} {\bibfnamefont {S.}~\bibnamefont
  {Pyon}}, \bibinfo {author} {\bibfnamefont {T.}~\bibnamefont {Takayama}},
  \bibinfo {author} {\bibfnamefont {H.}~\bibnamefont {Takagi}},\ and\ \bibinfo
  {author} {\bibfnamefont {A.}~\bibnamefont {Cavalleri}},\ }\href@noop {}
  {\bibfield  {journal} {\bibinfo  {journal} {Science}\ }\textbf {\bibinfo
  {volume} {331}},\ \bibinfo {pages} {189} (\bibinfo {year}
  {2011})}\BibitemShut {NoStop}%
\bibitem [{\citenamefont {Nova}\ \emph {et~al.}(2017)\citenamefont {Nova},
  \citenamefont {Cartella}, \citenamefont {Cantaluppi}, \citenamefont {Först},
  \citenamefont {Bossini}, \citenamefont {Mikhaylovskiy}, \citenamefont
  {Kimel}, \citenamefont {Merlin},\ and\ \citenamefont {Cavalleri}}]{Nova2017}%
  \BibitemOpen
  \bibfield  {author} {\bibinfo {author} {\bibfnamefont {T.~F.}\ \bibnamefont
  {Nova}}, \bibinfo {author} {\bibfnamefont {A.}~\bibnamefont {Cartella}},
  \bibinfo {author} {\bibfnamefont {A.}~\bibnamefont {Cantaluppi}}, \bibinfo
  {author} {\bibfnamefont {M.}~\bibnamefont {Först}}, \bibinfo {author}
  {\bibfnamefont {D.}~\bibnamefont {Bossini}}, \bibinfo {author} {\bibfnamefont
  {R.~V.}\ \bibnamefont {Mikhaylovskiy}}, \bibinfo {author} {\bibfnamefont
  {A.~V.}\ \bibnamefont {Kimel}}, \bibinfo {author} {\bibfnamefont
  {R.}~\bibnamefont {Merlin}},\ and\ \bibinfo {author} {\bibfnamefont
  {A.}~\bibnamefont {Cavalleri}},\ }\href@noop {} {\bibfield  {journal}
  {\bibinfo  {journal} {Nat. Phys.}\ }\textbf {\bibinfo {volume} {13}},\
  \bibinfo {pages} {132} (\bibinfo {year} {2017})}\BibitemShut {NoStop}%
\bibitem [{\citenamefont {Disa}\ \emph {et~al.}(2020)\citenamefont {Disa},
  \citenamefont {Fechner}, \citenamefont {Nova}, \citenamefont {Liu},
  \citenamefont {Först}, \citenamefont {Prabhakaran}, \citenamefont
  {Radaelli},\ and\ \citenamefont {Cavalleri}}]{Disa2020}%
  \BibitemOpen
  \bibfield  {author} {\bibinfo {author} {\bibfnamefont {A.~S.}\ \bibnamefont
  {Disa}}, \bibinfo {author} {\bibfnamefont {M.}~\bibnamefont {Fechner}},
  \bibinfo {author} {\bibfnamefont {T.~F.}\ \bibnamefont {Nova}}, \bibinfo
  {author} {\bibfnamefont {B.}~\bibnamefont {Liu}}, \bibinfo {author}
  {\bibfnamefont {M.}~\bibnamefont {Först}}, \bibinfo {author} {\bibfnamefont
  {D.}~\bibnamefont {Prabhakaran}}, \bibinfo {author} {\bibfnamefont {P.~G.}\
  \bibnamefont {Radaelli}},\ and\ \bibinfo {author} {\bibfnamefont
  {A.}~\bibnamefont {Cavalleri}},\ }\href@noop {} {\bibfield  {journal}
  {\bibinfo  {journal} {Nat. Phys.}\ }\textbf {\bibinfo {volume} {16}},\
  \bibinfo {pages} {937} (\bibinfo {year} {2020})}\BibitemShut {NoStop}%
\bibitem [{\citenamefont {Li}\ \emph {et~al.}(2019)\citenamefont {Li},
  \citenamefont {Qiu}, \citenamefont {Zhang}, \citenamefont {Baldini},
  \citenamefont {Lu}, \citenamefont {Rappe},\ and\ \citenamefont
  {Nelson}}]{Li2019}%
  \BibitemOpen
  \bibfield  {author} {\bibinfo {author} {\bibfnamefont {X.}~\bibnamefont
  {Li}}, \bibinfo {author} {\bibfnamefont {T.}~\bibnamefont {Qiu}}, \bibinfo
  {author} {\bibfnamefont {J.}~\bibnamefont {Zhang}}, \bibinfo {author}
  {\bibfnamefont {E.}~\bibnamefont {Baldini}}, \bibinfo {author} {\bibfnamefont
  {J.}~\bibnamefont {Lu}}, \bibinfo {author} {\bibfnamefont {A.~M.}\
  \bibnamefont {Rappe}},\ and\ \bibinfo {author} {\bibfnamefont {K.~A.}\
  \bibnamefont {Nelson}},\ }\href@noop {} {\bibfield  {journal} {\bibinfo
  {journal} {Science}\ }\textbf {\bibinfo {volume} {364}},\ \bibinfo {pages}
  {1079} (\bibinfo {year} {2019})}\BibitemShut {NoStop}%
\bibitem [{\citenamefont {Nova}\ \emph {et~al.}(2019)\citenamefont {Nova},
  \citenamefont {Disa}, \citenamefont {Fechner},\ and\ \citenamefont
  {Cavalleri}}]{Nova2019}%
  \BibitemOpen
  \bibfield  {author} {\bibinfo {author} {\bibfnamefont {T.~F.}\ \bibnamefont
  {Nova}}, \bibinfo {author} {\bibfnamefont {A.~S.}\ \bibnamefont {Disa}},
  \bibinfo {author} {\bibfnamefont {M.}~\bibnamefont {Fechner}},\ and\ \bibinfo
  {author} {\bibfnamefont {A.}~\bibnamefont {Cavalleri}},\ }\href@noop {}
  {\bibfield  {journal} {\bibinfo  {journal} {Science}\ }\textbf {\bibinfo
  {volume} {364}},\ \bibinfo {pages} {1075} (\bibinfo {year}
  {2019})}\BibitemShut {NoStop}%
\bibitem [{\citenamefont {Shin}\ \emph {et~al.}(2022)\citenamefont {Shin},
  \citenamefont {Latini}, \citenamefont {Schafer}, \citenamefont {Sato},
  \citenamefont {Baldini}, \citenamefont {Giovannini}, \citenamefont
  {Hubener},\ and\ \citenamefont {Rubio}}]{Shin2021arxiv}%
  \BibitemOpen
  \bibfield  {author} {\bibinfo {author} {\bibfnamefont {D.}~\bibnamefont
  {Shin}}, \bibinfo {author} {\bibfnamefont {S.}~\bibnamefont {Latini}},
  \bibinfo {author} {\bibfnamefont {C.}~\bibnamefont {Schafer}}, \bibinfo
  {author} {\bibfnamefont {S.~A.}\ \bibnamefont {Sato}}, \bibinfo {author}
  {\bibfnamefont {E.}~\bibnamefont {Baldini}}, \bibinfo {author} {\bibfnamefont
  {U.~D.}\ \bibnamefont {Giovannini}}, \bibinfo {author} {\bibfnamefont
  {H.}~\bibnamefont {Hubener}},\ and\ \bibinfo {author} {\bibfnamefont
  {A.}~\bibnamefont {Rubio}},\ }\href@noop {} {\bibfield  {journal} {\bibinfo
  {journal} {Phys. Rev. Lett.}\ }\textbf {\bibinfo {volume} {129}},\ \bibinfo
  {pages} {167401} (\bibinfo {year} {2022})}\BibitemShut {NoStop}%
\bibitem [{\citenamefont {Latini}\ \emph {et~al.}(2021)\citenamefont {Latini},
  \citenamefont {Shin}, \citenamefont {Sato}, \citenamefont {Schäfer},
  \citenamefont {Giovannini}, \citenamefont {Hübener},\ and\ \citenamefont
  {Rubio}}]{Simone2021}%
  \BibitemOpen
  \bibfield  {author} {\bibinfo {author} {\bibfnamefont {S.}~\bibnamefont
  {Latini}}, \bibinfo {author} {\bibfnamefont {D.}~\bibnamefont {Shin}},
  \bibinfo {author} {\bibfnamefont {S.~A.}\ \bibnamefont {Sato}}, \bibinfo
  {author} {\bibfnamefont {C.}~\bibnamefont {Schäfer}}, \bibinfo {author}
  {\bibfnamefont {U.~D.}\ \bibnamefont {Giovannini}}, \bibinfo {author}
  {\bibfnamefont {H.}~\bibnamefont {Hübener}},\ and\ \bibinfo {author}
  {\bibfnamefont {A.}~\bibnamefont {Rubio}},\ }\href@noop {} {\bibfield
  {journal} {\bibinfo  {journal} {Proc. Natl. Acad. Sci. U.S.A.}\ }\textbf
  {\bibinfo {volume} {118}},\ \bibinfo {pages} {e2105618118} (\bibinfo {year}
  {2021})}\BibitemShut {NoStop}%
\bibitem [{\citenamefont {Pueyo}\ and\ \citenamefont
  {Subedi}(2022)}]{pueyo2022}%
  \BibitemOpen
  \bibfield  {author} {\bibinfo {author} {\bibfnamefont {A.~G.}\ \bibnamefont
  {Pueyo}}\ and\ \bibinfo {author} {\bibfnamefont {A.}~\bibnamefont {Subedi}},\
  }\href {https://doi.org/10.1103/physrevb.106.214305} {\bibfield  {journal}
  {\bibinfo  {journal} {Phys. Rev. B}\ }\textbf {\bibinfo {volume} {106}},\
  \bibinfo {pages} {214305} (\bibinfo {year} {2022})},\ \Eprint
  {https://arxiv.org/abs/2208.05293} {2208.05293} \BibitemShut {NoStop}%
\bibitem [{\citenamefont {Subedi}\ \emph {et~al.}(2014)\citenamefont {Subedi},
  \citenamefont {Cavalleri},\ and\ \citenamefont {Georges}}]{subedi2014}%
  \BibitemOpen
  \bibfield  {author} {\bibinfo {author} {\bibfnamefont {A.}~\bibnamefont
  {Subedi}}, \bibinfo {author} {\bibfnamefont {A.}~\bibnamefont {Cavalleri}},\
  and\ \bibinfo {author} {\bibfnamefont {A.}~\bibnamefont {Georges}},\
  }\href@noop {} {\bibfield  {journal} {\bibinfo  {journal} {Phys. Rev. B}\
  }\textbf {\bibinfo {volume} {89}},\ \bibinfo {pages} {220301} (\bibinfo
  {year} {2014})}\BibitemShut {NoStop}%
\bibitem [{\citenamefont {Armitage}\ \emph {et~al.}(2018)\citenamefont
  {Armitage}, \citenamefont {Mele},\ and\ \citenamefont
  {Vishwanath}}]{Armitage2018}%
  \BibitemOpen
  \bibfield  {author} {\bibinfo {author} {\bibfnamefont {N.~P.}\ \bibnamefont
  {Armitage}}, \bibinfo {author} {\bibfnamefont {E.~J.}\ \bibnamefont {Mele}},\
  and\ \bibinfo {author} {\bibfnamefont {A.}~\bibnamefont {Vishwanath}},\
  }\href {https://doi.org/10.1103/RevModPhys.90.015001} {\bibfield  {journal}
  {\bibinfo  {journal} {Rev. Mod. Phys.}\ }\textbf {\bibinfo {volume} {90}},\
  \bibinfo {pages} {015001} (\bibinfo {year} {2018})}\BibitemShut {NoStop}%
\bibitem [{\citenamefont {Liu}\ \emph {et~al.}(2020)\citenamefont {Liu},
  \citenamefont {Xia}, \citenamefont {Xiao}, \citenamefont {Garcia~de Abajo},\
  and\ \citenamefont {Sun}}]{liu2020semimetals}%
  \BibitemOpen
  \bibfield  {author} {\bibinfo {author} {\bibfnamefont {J.}~\bibnamefont
  {Liu}}, \bibinfo {author} {\bibfnamefont {F.}~\bibnamefont {Xia}}, \bibinfo
  {author} {\bibfnamefont {D.}~\bibnamefont {Xiao}}, \bibinfo {author}
  {\bibfnamefont {F.~J.}\ \bibnamefont {Garcia~de Abajo}},\ and\ \bibinfo
  {author} {\bibfnamefont {D.}~\bibnamefont {Sun}},\ }\href@noop {} {\bibfield
  {journal} {\bibinfo  {journal} {Nat. Mater.}\ }\textbf {\bibinfo {volume}
  {19}},\ \bibinfo {pages} {830} (\bibinfo {year} {2020})}\BibitemShut
  {NoStop}%
\bibitem [{\citenamefont {McIver}\ \emph {et~al.}(2020)\citenamefont {McIver},
  \citenamefont {Schulte}, \citenamefont {Stein}, \citenamefont {Matsuyama},
  \citenamefont {Jotzu}, \citenamefont {Meier},\ and\ \citenamefont
  {Cavalleri}}]{McIver2020}%
  \BibitemOpen
  \bibfield  {author} {\bibinfo {author} {\bibfnamefont {J.~W.}\ \bibnamefont
  {McIver}}, \bibinfo {author} {\bibfnamefont {B.}~\bibnamefont {Schulte}},
  \bibinfo {author} {\bibfnamefont {F.-U.}\ \bibnamefont {Stein}}, \bibinfo
  {author} {\bibfnamefont {T.}~\bibnamefont {Matsuyama}}, \bibinfo {author}
  {\bibfnamefont {G.}~\bibnamefont {Jotzu}}, \bibinfo {author} {\bibfnamefont
  {G.}~\bibnamefont {Meier}},\ and\ \bibinfo {author} {\bibfnamefont
  {A.}~\bibnamefont {Cavalleri}},\ }\href@noop {} {\bibfield  {journal}
  {\bibinfo  {journal} {Nat. Phys.}\ }\textbf {\bibinfo {volume} {16}},\
  \bibinfo {pages} {38} (\bibinfo {year} {2020})}\BibitemShut {NoStop}%
\bibitem [{\citenamefont {Sato}\ \emph {et~al.}(2019)\citenamefont {Sato},
  \citenamefont {McIver}, \citenamefont {Nuske}, \citenamefont {Tang},
  \citenamefont {Jotzu}, \citenamefont {Schulte}, \citenamefont {Hübener},
  \citenamefont {Giovannini}, \citenamefont {Mathey}, \citenamefont {Sentef},
  \citenamefont {Cavalleri},\ and\ \citenamefont {Rubio}}]{Sato2019}%
  \BibitemOpen
  \bibfield  {author} {\bibinfo {author} {\bibfnamefont {S.~A.}\ \bibnamefont
  {Sato}}, \bibinfo {author} {\bibfnamefont {J.~W.}\ \bibnamefont {McIver}},
  \bibinfo {author} {\bibfnamefont {M.}~\bibnamefont {Nuske}}, \bibinfo
  {author} {\bibfnamefont {P.}~\bibnamefont {Tang}}, \bibinfo {author}
  {\bibfnamefont {G.}~\bibnamefont {Jotzu}}, \bibinfo {author} {\bibfnamefont
  {B.}~\bibnamefont {Schulte}}, \bibinfo {author} {\bibfnamefont
  {H.}~\bibnamefont {Hübener}}, \bibinfo {author} {\bibfnamefont {U.~D.}\
  \bibnamefont {Giovannini}}, \bibinfo {author} {\bibfnamefont
  {L.}~\bibnamefont {Mathey}}, \bibinfo {author} {\bibfnamefont {M.~A.}\
  \bibnamefont {Sentef}}, \bibinfo {author} {\bibfnamefont {A.}~\bibnamefont
  {Cavalleri}},\ and\ \bibinfo {author} {\bibfnamefont {A.}~\bibnamefont
  {Rubio}},\ }\href@noop {} {\bibfield  {journal} {\bibinfo  {journal} {Phys.
  Rev. B}\ }\textbf {\bibinfo {volume} {99}},\ \bibinfo {pages} {214302}
  (\bibinfo {year} {2019})}\BibitemShut {NoStop}%
\bibitem [{\citenamefont {H{\"u}bener}\ \emph {et~al.}(2017)\citenamefont
  {H{\"u}bener}, \citenamefont {Sentef}, \citenamefont {De~Giovannini},
  \citenamefont {Kemper},\ and\ \citenamefont {Rubio}}]{hubener2017creating}%
  \BibitemOpen
  \bibfield  {author} {\bibinfo {author} {\bibfnamefont {H.}~\bibnamefont
  {H{\"u}bener}}, \bibinfo {author} {\bibfnamefont {M.~A.}\ \bibnamefont
  {Sentef}}, \bibinfo {author} {\bibfnamefont {U.}~\bibnamefont
  {De~Giovannini}}, \bibinfo {author} {\bibfnamefont {A.~F.}\ \bibnamefont
  {Kemper}},\ and\ \bibinfo {author} {\bibfnamefont {A.}~\bibnamefont
  {Rubio}},\ }\href@noop {} {\bibfield  {journal} {\bibinfo  {journal} {Nat.
  Commun.}\ }\textbf {\bibinfo {volume} {8}},\ \bibinfo {pages} {1} (\bibinfo
  {year} {2017})}\BibitemShut {NoStop}%
\bibitem [{\citenamefont {Sie}\ \emph {et~al.}(2019)\citenamefont {Sie},
  \citenamefont {Nyby}, \citenamefont {Pemmaraju}, \citenamefont {Park},
  \citenamefont {Shen}, \citenamefont {Yang}, \citenamefont {Hoffmann},
  \citenamefont {Ofori-Okai}, \citenamefont {Li}, \citenamefont {Reid},
  \citenamefont {Weathersby}, \citenamefont {Mannebach}, \citenamefont
  {Finney}, \citenamefont {Rhodes}, \citenamefont {Chenet}, \citenamefont
  {Antony}, \citenamefont {Balicas}, \citenamefont {Hone}, \citenamefont
  {Devereaux}, \citenamefont {Heinz}, \citenamefont {Wang},\ and\ \citenamefont
  {Lindenberg}}]{Sie2019}%
  \BibitemOpen
  \bibfield  {author} {\bibinfo {author} {\bibfnamefont {E.~J.}\ \bibnamefont
  {Sie}}, \bibinfo {author} {\bibfnamefont {C.~M.}\ \bibnamefont {Nyby}},
  \bibinfo {author} {\bibfnamefont {C.~D.}\ \bibnamefont {Pemmaraju}}, \bibinfo
  {author} {\bibfnamefont {S.~J.}\ \bibnamefont {Park}}, \bibinfo {author}
  {\bibfnamefont {X.}~\bibnamefont {Shen}}, \bibinfo {author} {\bibfnamefont
  {J.}~\bibnamefont {Yang}}, \bibinfo {author} {\bibfnamefont {M.~C.}\
  \bibnamefont {Hoffmann}}, \bibinfo {author} {\bibfnamefont {B.~K.}\
  \bibnamefont {Ofori-Okai}}, \bibinfo {author} {\bibfnamefont
  {R.}~\bibnamefont {Li}}, \bibinfo {author} {\bibfnamefont {A.~H.}\
  \bibnamefont {Reid}}, \bibinfo {author} {\bibfnamefont {S.}~\bibnamefont
  {Weathersby}}, \bibinfo {author} {\bibfnamefont {E.}~\bibnamefont
  {Mannebach}}, \bibinfo {author} {\bibfnamefont {N.}~\bibnamefont {Finney}},
  \bibinfo {author} {\bibfnamefont {D.}~\bibnamefont {Rhodes}}, \bibinfo
  {author} {\bibfnamefont {D.}~\bibnamefont {Chenet}}, \bibinfo {author}
  {\bibfnamefont {A.}~\bibnamefont {Antony}}, \bibinfo {author} {\bibfnamefont
  {L.}~\bibnamefont {Balicas}}, \bibinfo {author} {\bibfnamefont
  {J.}~\bibnamefont {Hone}}, \bibinfo {author} {\bibfnamefont {T.~P.}\
  \bibnamefont {Devereaux}}, \bibinfo {author} {\bibfnamefont {T.~F.}\
  \bibnamefont {Heinz}}, \bibinfo {author} {\bibfnamefont {X.}~\bibnamefont
  {Wang}},\ and\ \bibinfo {author} {\bibfnamefont {A.~M.}\ \bibnamefont
  {Lindenberg}},\ }\href@noop {} {\bibfield  {journal} {\bibinfo  {journal}
  {Nature}\ }\textbf {\bibinfo {volume} {565}},\ \bibinfo {pages} {61}
  (\bibinfo {year} {2019})}\BibitemShut {NoStop}%
\bibitem [{\citenamefont {Vaswani}\ \emph {et~al.}(2020)\citenamefont
  {Vaswani}, \citenamefont {Wang}, \citenamefont {Mudiyanselage}, \citenamefont
  {Li}, \citenamefont {Lozano}, \citenamefont {Gu}, \citenamefont {Cheng},
  \citenamefont {Song}, \citenamefont {Luo}, \citenamefont {Kim}, \citenamefont
  {Huang}, \citenamefont {Liu}, \citenamefont {Mootz}, \citenamefont {Perakis},
  \citenamefont {Yao}, \citenamefont {Ho},\ and\ \citenamefont
  {Wang}}]{Vaswani2020}%
  \BibitemOpen
  \bibfield  {author} {\bibinfo {author} {\bibfnamefont {C.}~\bibnamefont
  {Vaswani}}, \bibinfo {author} {\bibfnamefont {L.-L.}\ \bibnamefont {Wang}},
  \bibinfo {author} {\bibfnamefont {D.~H.}\ \bibnamefont {Mudiyanselage}},
  \bibinfo {author} {\bibfnamefont {Q.}~\bibnamefont {Li}}, \bibinfo {author}
  {\bibfnamefont {P.~M.}\ \bibnamefont {Lozano}}, \bibinfo {author}
  {\bibfnamefont {G.~D.}\ \bibnamefont {Gu}}, \bibinfo {author} {\bibfnamefont
  {D.}~\bibnamefont {Cheng}}, \bibinfo {author} {\bibfnamefont
  {B.}~\bibnamefont {Song}}, \bibinfo {author} {\bibfnamefont {L.}~\bibnamefont
  {Luo}}, \bibinfo {author} {\bibfnamefont {R.~H.~J.}\ \bibnamefont {Kim}},
  \bibinfo {author} {\bibfnamefont {C.}~\bibnamefont {Huang}}, \bibinfo
  {author} {\bibfnamefont {Z.}~\bibnamefont {Liu}}, \bibinfo {author}
  {\bibfnamefont {M.}~\bibnamefont {Mootz}}, \bibinfo {author} {\bibfnamefont
  {I.~E.}\ \bibnamefont {Perakis}}, \bibinfo {author} {\bibfnamefont
  {Y.}~\bibnamefont {Yao}}, \bibinfo {author} {\bibfnamefont {K.~M.}\
  \bibnamefont {Ho}},\ and\ \bibinfo {author} {\bibfnamefont {J.}~\bibnamefont
  {Wang}},\ }\href@noop {} {\bibfield  {journal} {\bibinfo  {journal} {Phys.
  Rev. X}\ }\textbf {\bibinfo {volume} {10}},\ \bibinfo {pages} {021013}
  (\bibinfo {year} {2020})}\BibitemShut {NoStop}%
\bibitem [{\citenamefont {Guan}\ \emph {et~al.}(2021)\citenamefont {Guan},
  \citenamefont {Wang}, \citenamefont {You}, \citenamefont {Sun},\ and\
  \citenamefont {Meng}}]{guan2021manipulating}%
  \BibitemOpen
  \bibfield  {author} {\bibinfo {author} {\bibfnamefont {M.-X.}\ \bibnamefont
  {Guan}}, \bibinfo {author} {\bibfnamefont {E.}~\bibnamefont {Wang}}, \bibinfo
  {author} {\bibfnamefont {P.-W.}\ \bibnamefont {You}}, \bibinfo {author}
  {\bibfnamefont {J.-T.}\ \bibnamefont {Sun}},\ and\ \bibinfo {author}
  {\bibfnamefont {S.}~\bibnamefont {Meng}},\ }\href@noop {} {\bibfield
  {journal} {\bibinfo  {journal} {Nat. Commun.}\ }\textbf {\bibinfo {volume}
  {12}},\ \bibinfo {pages} {1} (\bibinfo {year} {2021})}\BibitemShut {NoStop}%
\bibitem [{\citenamefont {Luo}\ \emph {et~al.}(2021)\citenamefont {Luo},
  \citenamefont {Cheng}, \citenamefont {Song}, \citenamefont {Wang},
  \citenamefont {Vaswani}, \citenamefont {Lozano}, \citenamefont {Gu},
  \citenamefont {Huang}, \citenamefont {Kim}, \citenamefont {Liu},
  \citenamefont {Park}, \citenamefont {Yao}, \citenamefont {Ho}, \citenamefont
  {Perakis}, \citenamefont {Li},\ and\ \citenamefont {Wang}}]{Luo2021}%
  \BibitemOpen
  \bibfield  {author} {\bibinfo {author} {\bibfnamefont {L.}~\bibnamefont
  {Luo}}, \bibinfo {author} {\bibfnamefont {D.}~\bibnamefont {Cheng}}, \bibinfo
  {author} {\bibfnamefont {B.}~\bibnamefont {Song}}, \bibinfo {author}
  {\bibfnamefont {L.-L.}\ \bibnamefont {Wang}}, \bibinfo {author}
  {\bibfnamefont {C.}~\bibnamefont {Vaswani}}, \bibinfo {author} {\bibfnamefont
  {P.~M.}\ \bibnamefont {Lozano}}, \bibinfo {author} {\bibfnamefont
  {G.}~\bibnamefont {Gu}}, \bibinfo {author} {\bibfnamefont {C.}~\bibnamefont
  {Huang}}, \bibinfo {author} {\bibfnamefont {R.~H.~J.}\ \bibnamefont {Kim}},
  \bibinfo {author} {\bibfnamefont {Z.}~\bibnamefont {Liu}}, \bibinfo {author}
  {\bibfnamefont {J.-M.}\ \bibnamefont {Park}}, \bibinfo {author}
  {\bibfnamefont {Y.}~\bibnamefont {Yao}}, \bibinfo {author} {\bibfnamefont
  {K.}~\bibnamefont {Ho}}, \bibinfo {author} {\bibfnamefont {I.~E.}\
  \bibnamefont {Perakis}}, \bibinfo {author} {\bibfnamefont {Q.}~\bibnamefont
  {Li}},\ and\ \bibinfo {author} {\bibfnamefont {J.}~\bibnamefont {Wang}},\
  }\href@noop {} {\bibfield  {journal} {\bibinfo  {journal} {Nat. Mater.}\
  }\textbf {\bibinfo {volume} {20}},\ \bibinfo {pages} {329} (\bibinfo {year}
  {2021})}\BibitemShut {NoStop}%
\bibitem [{\citenamefont {Hu}\ \emph {et~al.}(2023)\citenamefont {Hu},
  \citenamefont {Su}, \citenamefont {Shi}, \citenamefont {Wang}, \citenamefont
  {Yue}, \citenamefont {Xu}, \citenamefont {Zhang}, \citenamefont {Liu},
  \citenamefont {Wu}, \citenamefont {Li}, \citenamefont {Zhou}, \citenamefont
  {Yuan}, \citenamefont {Wu}, \citenamefont {Chen}, \citenamefont {Dong},\ and\
  \citenamefont {Wang}}]{Hu2023}%
  \BibitemOpen
  \bibfield  {author} {\bibinfo {author} {\bibfnamefont {T.}~\bibnamefont
  {Hu}}, \bibinfo {author} {\bibfnamefont {B.}~\bibnamefont {Su}}, \bibinfo
  {author} {\bibfnamefont {L.}~\bibnamefont {Shi}}, \bibinfo {author}
  {\bibfnamefont {Z.}~\bibnamefont {Wang}}, \bibinfo {author} {\bibfnamefont
  {L.}~\bibnamefont {Yue}}, \bibinfo {author} {\bibfnamefont {S.}~\bibnamefont
  {Xu}}, \bibinfo {author} {\bibfnamefont {S.}~\bibnamefont {Zhang}}, \bibinfo
  {author} {\bibfnamefont {Q.}~\bibnamefont {Liu}}, \bibinfo {author}
  {\bibfnamefont {Q.}~\bibnamefont {Wu}}, \bibinfo {author} {\bibfnamefont
  {R.}~\bibnamefont {Li}}, \bibinfo {author} {\bibfnamefont {X.}~\bibnamefont
  {Zhou}}, \bibinfo {author} {\bibfnamefont {J.}~\bibnamefont {Yuan}}, \bibinfo
  {author} {\bibfnamefont {D.}~\bibnamefont {Wu}}, \bibinfo {author}
  {\bibfnamefont {Z.}~\bibnamefont {Chen}}, \bibinfo {author} {\bibfnamefont
  {T.}~\bibnamefont {Dong}},\ and\ \bibinfo {author} {\bibfnamefont
  {N.}~\bibnamefont {Wang}},\ }\href@noop {} {\bibfield  {journal} {\bibinfo
  {journal} {Adv. Opt. Mater.}\ }\textbf {\bibinfo {volume} {11}},\ \bibinfo
  {pages} {2202639} (\bibinfo {year} {2023})}\BibitemShut {NoStop}%
\bibitem [{\citenamefont {Ning}\ \emph {et~al.}(2022)\citenamefont {Ning},
  \citenamefont {Mehio}, \citenamefont {Lian}, \citenamefont {Li},
  \citenamefont {Zoghlin}, \citenamefont {Zhou}, \citenamefont {Cheng},
  \citenamefont {Wilson}, \citenamefont {Wong},\ and\ \citenamefont
  {Hsieh}}]{Ning2022}%
  \BibitemOpen
  \bibfield  {author} {\bibinfo {author} {\bibfnamefont {H.}~\bibnamefont
  {Ning}}, \bibinfo {author} {\bibfnamefont {O.}~\bibnamefont {Mehio}},
  \bibinfo {author} {\bibfnamefont {C.}~\bibnamefont {Lian}}, \bibinfo {author}
  {\bibfnamefont {X.}~\bibnamefont {Li}}, \bibinfo {author} {\bibfnamefont
  {E.}~\bibnamefont {Zoghlin}}, \bibinfo {author} {\bibfnamefont
  {P.}~\bibnamefont {Zhou}}, \bibinfo {author} {\bibfnamefont {B.}~\bibnamefont
  {Cheng}}, \bibinfo {author} {\bibfnamefont {S.~D.}\ \bibnamefont {Wilson}},
  \bibinfo {author} {\bibfnamefont {B.~M.}\ \bibnamefont {Wong}},\ and\
  \bibinfo {author} {\bibfnamefont {D.}~\bibnamefont {Hsieh}},\ }\href@noop {}
  {\bibfield  {journal} {\bibinfo  {journal} {Phys. Rev. B}\ }\textbf {\bibinfo
  {volume} {106}},\ \bibinfo {pages} {205118} (\bibinfo {year}
  {2022})}\BibitemShut {NoStop}%
\bibitem [{\citenamefont {Jnawali}\ \emph {et~al.}(2020)\citenamefont
  {Jnawali}, \citenamefont {Xiang}, \citenamefont {Linser}, \citenamefont
  {Shojaei}, \citenamefont {Wang}, \citenamefont {Qiu}, \citenamefont {Lian},
  \citenamefont {Wong}, \citenamefont {Wu}, \citenamefont {Ye}, \citenamefont
  {Leng}, \citenamefont {Jackson},\ and\ \citenamefont {Smith}}]{Jnawali2020}%
  \BibitemOpen
  \bibfield  {author} {\bibinfo {author} {\bibfnamefont {G.}~\bibnamefont
  {Jnawali}}, \bibinfo {author} {\bibfnamefont {Y.}~\bibnamefont {Xiang}},
  \bibinfo {author} {\bibfnamefont {S.~M.}\ \bibnamefont {Linser}}, \bibinfo
  {author} {\bibfnamefont {I.~A.}\ \bibnamefont {Shojaei}}, \bibinfo {author}
  {\bibfnamefont {R.}~\bibnamefont {Wang}}, \bibinfo {author} {\bibfnamefont
  {G.}~\bibnamefont {Qiu}}, \bibinfo {author} {\bibfnamefont {C.}~\bibnamefont
  {Lian}}, \bibinfo {author} {\bibfnamefont {B.~M.}\ \bibnamefont {Wong}},
  \bibinfo {author} {\bibfnamefont {W.}~\bibnamefont {Wu}}, \bibinfo {author}
  {\bibfnamefont {P.~D.}\ \bibnamefont {Ye}}, \bibinfo {author} {\bibfnamefont
  {Y.}~\bibnamefont {Leng}}, \bibinfo {author} {\bibfnamefont {H.~E.}\
  \bibnamefont {Jackson}},\ and\ \bibinfo {author} {\bibfnamefont {L.~M.}\
  \bibnamefont {Smith}},\ }\href@noop {} {\bibfield  {journal} {\bibinfo
  {journal} {Nat. Commun.}\ }\textbf {\bibinfo {volume} {11}},\ \bibinfo
  {pages} {3991} (\bibinfo {year} {2020})}\BibitemShut {NoStop}%
\bibitem [{\citenamefont {Deng}\ \emph {et~al.}(2016)\citenamefont {Deng},
  \citenamefont {Wan}, \citenamefont {Deng}, \citenamefont {Zhang},
  \citenamefont {Ding}, \citenamefont {Wang}, \citenamefont {Yan},
  \citenamefont {Huang}, \citenamefont {Zhang}, \citenamefont {Xu},
  \citenamefont {Denlinger}, \citenamefont {Fedorov}, \citenamefont {Yang},
  \citenamefont {Duan}, \citenamefont {Yao}, \citenamefont {Wu}, \citenamefont
  {Fan}, \citenamefont {Zhang}, \citenamefont {Chen},\ and\ \citenamefont
  {Zhou}}]{Deng2016}%
  \BibitemOpen
  \bibfield  {author} {\bibinfo {author} {\bibfnamefont {K.}~\bibnamefont
  {Deng}}, \bibinfo {author} {\bibfnamefont {G.}~\bibnamefont {Wan}}, \bibinfo
  {author} {\bibfnamefont {P.}~\bibnamefont {Deng}}, \bibinfo {author}
  {\bibfnamefont {K.}~\bibnamefont {Zhang}}, \bibinfo {author} {\bibfnamefont
  {S.}~\bibnamefont {Ding}}, \bibinfo {author} {\bibfnamefont {E.}~\bibnamefont
  {Wang}}, \bibinfo {author} {\bibfnamefont {M.}~\bibnamefont {Yan}}, \bibinfo
  {author} {\bibfnamefont {H.}~\bibnamefont {Huang}}, \bibinfo {author}
  {\bibfnamefont {H.}~\bibnamefont {Zhang}}, \bibinfo {author} {\bibfnamefont
  {Z.}~\bibnamefont {Xu}}, \bibinfo {author} {\bibfnamefont {J.}~\bibnamefont
  {Denlinger}}, \bibinfo {author} {\bibfnamefont {A.}~\bibnamefont {Fedorov}},
  \bibinfo {author} {\bibfnamefont {H.}~\bibnamefont {Yang}}, \bibinfo {author}
  {\bibfnamefont {W.}~\bibnamefont {Duan}}, \bibinfo {author} {\bibfnamefont
  {H.}~\bibnamefont {Yao}}, \bibinfo {author} {\bibfnamefont {Y.}~\bibnamefont
  {Wu}}, \bibinfo {author} {\bibfnamefont {S.}~\bibnamefont {Fan}}, \bibinfo
  {author} {\bibfnamefont {H.}~\bibnamefont {Zhang}}, \bibinfo {author}
  {\bibfnamefont {X.}~\bibnamefont {Chen}},\ and\ \bibinfo {author}
  {\bibfnamefont {S.}~\bibnamefont {Zhou}},\ }\href@noop {} {\bibfield
  {journal} {\bibinfo  {journal} {Nat. Phys.}\ }\textbf {\bibinfo {volume}
  {12}},\ \bibinfo {pages} {1105} (\bibinfo {year} {2016})}\BibitemShut
  {NoStop}%
\bibitem [{\citenamefont {Lu}\ \emph {et~al.}(2022)\citenamefont {Lu},
  \citenamefont {Zhou}, \citenamefont {Park},\ and\ \citenamefont
  {Bernardi}}]{Lu2022}%
  \BibitemOpen
  \bibfield  {author} {\bibinfo {author} {\bibfnamefont {I.-T.}\ \bibnamefont
  {Lu}}, \bibinfo {author} {\bibfnamefont {J.-J.}\ \bibnamefont {Zhou}},
  \bibinfo {author} {\bibfnamefont {J.}~\bibnamefont {Park}},\ and\ \bibinfo
  {author} {\bibfnamefont {M.}~\bibnamefont {Bernardi}},\ }\href@noop {}
  {\bibfield  {journal} {\bibinfo  {journal} {Phys. Rev. Mater.}\ }\textbf
  {\bibinfo {volume} {6}},\ \bibinfo {pages} {L010801} (\bibinfo {year}
  {2022})}\BibitemShut {NoStop}%
\bibitem [{\citenamefont {Malic}\ \emph {et~al.}(2011)\citenamefont {Malic},
  \citenamefont {Winzer}, \citenamefont {Bobkin},\ and\ \citenamefont
  {Knorr}}]{Malic2011}%
  \BibitemOpen
  \bibfield  {author} {\bibinfo {author} {\bibfnamefont {E.}~\bibnamefont
  {Malic}}, \bibinfo {author} {\bibfnamefont {T.}~\bibnamefont {Winzer}},
  \bibinfo {author} {\bibfnamefont {E.}~\bibnamefont {Bobkin}},\ and\ \bibinfo
  {author} {\bibfnamefont {A.}~\bibnamefont {Knorr}},\ }\href
  {https://doi.org/10.1103/PhysRevB.84.205406} {\bibfield  {journal} {\bibinfo
  {journal} {Phys. Rev. B}\ }\textbf {\bibinfo {volume} {84}},\ \bibinfo
  {pages} {205406} (\bibinfo {year} {2011})}\BibitemShut {NoStop}%
\bibitem [{\citenamefont {Xu}\ \emph {et~al.}(2015)\citenamefont {Xu},
  \citenamefont {Liu}, \citenamefont {Kushwaha}, \citenamefont {Sankar},
  \citenamefont {Krizan}, \citenamefont {Belopolski}, \citenamefont {Neupane},
  \citenamefont {Bian}, \citenamefont {Alidoust}, \citenamefont {Chang},
  \citenamefont {Jeng}, \citenamefont {Huang}, \citenamefont {Tsai},
  \citenamefont {Lin}, \citenamefont {Shibayev}, \citenamefont {Chou},
  \citenamefont {Cava},\ and\ \citenamefont {Hasan}}]{Xu2015}%
  \BibitemOpen
  \bibfield  {author} {\bibinfo {author} {\bibfnamefont {S.-Y.}\ \bibnamefont
  {Xu}}, \bibinfo {author} {\bibfnamefont {C.}~\bibnamefont {Liu}}, \bibinfo
  {author} {\bibfnamefont {S.~K.}\ \bibnamefont {Kushwaha}}, \bibinfo {author}
  {\bibfnamefont {R.}~\bibnamefont {Sankar}}, \bibinfo {author} {\bibfnamefont
  {J.~W.}\ \bibnamefont {Krizan}}, \bibinfo {author} {\bibfnamefont
  {I.}~\bibnamefont {Belopolski}}, \bibinfo {author} {\bibfnamefont
  {M.}~\bibnamefont {Neupane}}, \bibinfo {author} {\bibfnamefont
  {G.}~\bibnamefont {Bian}}, \bibinfo {author} {\bibfnamefont {N.}~\bibnamefont
  {Alidoust}}, \bibinfo {author} {\bibfnamefont {T.-R.}\ \bibnamefont {Chang}},
  \bibinfo {author} {\bibfnamefont {H.-T.}\ \bibnamefont {Jeng}}, \bibinfo
  {author} {\bibfnamefont {C.-Y.}\ \bibnamefont {Huang}}, \bibinfo {author}
  {\bibfnamefont {W.-F.}\ \bibnamefont {Tsai}}, \bibinfo {author}
  {\bibfnamefont {H.}~\bibnamefont {Lin}}, \bibinfo {author} {\bibfnamefont
  {P.~P.}\ \bibnamefont {Shibayev}}, \bibinfo {author} {\bibfnamefont {F.-C.}\
  \bibnamefont {Chou}}, \bibinfo {author} {\bibfnamefont {R.~J.}\ \bibnamefont
  {Cava}},\ and\ \bibinfo {author} {\bibfnamefont {M.~Z.}\ \bibnamefont
  {Hasan}},\ }\href@noop {} {\bibfield  {journal} {\bibinfo  {journal}
  {Science}\ }\textbf {\bibinfo {volume} {347}},\ \bibinfo {pages} {294}
  (\bibinfo {year} {2015})}\BibitemShut {NoStop}%
\bibitem [{\citenamefont {Yang}\ \emph {et~al.}(2015)\citenamefont {Yang},
  \citenamefont {Liu}, \citenamefont {Sun}, \citenamefont {Peng}, \citenamefont
  {Yang}, \citenamefont {Zhang}, \citenamefont {Zhou}, \citenamefont {Zhang},
  \citenamefont {Guo}, \citenamefont {Rahn}, \citenamefont {Prabhakaran},
  \citenamefont {Hussain}, \citenamefont {Mo}, \citenamefont {Felser},
  \citenamefont {Yan},\ and\ \citenamefont {Chen}}]{Yang2015}%
  \BibitemOpen
  \bibfield  {author} {\bibinfo {author} {\bibfnamefont {L.~X.}\ \bibnamefont
  {Yang}}, \bibinfo {author} {\bibfnamefont {Z.~K.}\ \bibnamefont {Liu}},
  \bibinfo {author} {\bibfnamefont {Y.}~\bibnamefont {Sun}}, \bibinfo {author}
  {\bibfnamefont {H.}~\bibnamefont {Peng}}, \bibinfo {author} {\bibfnamefont
  {H.~F.}\ \bibnamefont {Yang}}, \bibinfo {author} {\bibfnamefont
  {T.}~\bibnamefont {Zhang}}, \bibinfo {author} {\bibfnamefont
  {B.}~\bibnamefont {Zhou}}, \bibinfo {author} {\bibfnamefont {Y.}~\bibnamefont
  {Zhang}}, \bibinfo {author} {\bibfnamefont {Y.~F.}\ \bibnamefont {Guo}},
  \bibinfo {author} {\bibfnamefont {M.}~\bibnamefont {Rahn}}, \bibinfo {author}
  {\bibfnamefont {D.}~\bibnamefont {Prabhakaran}}, \bibinfo {author}
  {\bibfnamefont {Z.}~\bibnamefont {Hussain}}, \bibinfo {author} {\bibfnamefont
  {S.-K.}\ \bibnamefont {Mo}}, \bibinfo {author} {\bibfnamefont
  {C.}~\bibnamefont {Felser}}, \bibinfo {author} {\bibfnamefont
  {B.}~\bibnamefont {Yan}},\ and\ \bibinfo {author} {\bibfnamefont {Y.~L.}\
  \bibnamefont {Chen}},\ }\href@noop {} {\bibfield  {journal} {\bibinfo
  {journal} {Nat. Phys.}\ }\textbf {\bibinfo {volume} {11}},\ \bibinfo {pages}
  {728} (\bibinfo {year} {2015})}\BibitemShut {NoStop}%
\bibitem [{\citenamefont {Gierz}\ \emph {et~al.}(2013)\citenamefont {Gierz},
  \citenamefont {Petersen}, \citenamefont {Mitrano}, \citenamefont {Cacho},
  \citenamefont {Turcu}, \citenamefont {Springate}, \citenamefont {Stöhr},
  \citenamefont {Köhler}, \citenamefont {Starke},\ and\ \citenamefont
  {Cavalleri}}]{Gierz2013}%
  \BibitemOpen
  \bibfield  {author} {\bibinfo {author} {\bibfnamefont {I.}~\bibnamefont
  {Gierz}}, \bibinfo {author} {\bibfnamefont {J.~C.}\ \bibnamefont {Petersen}},
  \bibinfo {author} {\bibfnamefont {M.}~\bibnamefont {Mitrano}}, \bibinfo
  {author} {\bibfnamefont {C.}~\bibnamefont {Cacho}}, \bibinfo {author}
  {\bibfnamefont {I.~C.~E.}\ \bibnamefont {Turcu}}, \bibinfo {author}
  {\bibfnamefont {E.}~\bibnamefont {Springate}}, \bibinfo {author}
  {\bibfnamefont {A.}~\bibnamefont {Stöhr}}, \bibinfo {author} {\bibfnamefont
  {A.}~\bibnamefont {Köhler}}, \bibinfo {author} {\bibfnamefont
  {U.}~\bibnamefont {Starke}},\ and\ \bibinfo {author} {\bibfnamefont
  {A.}~\bibnamefont {Cavalleri}},\ }\href@noop {} {\bibfield  {journal}
  {\bibinfo  {journal} {Nat. Mater.}\ }\textbf {\bibinfo {volume} {12}},\
  \bibinfo {pages} {1119} (\bibinfo {year} {2013})}\BibitemShut {NoStop}%
\bibitem [{SM()}]{SM}%
  \BibitemOpen
  \href@noop {} {\bibinfo {title} {See supplemental material at
  url}}\BibitemShut {NoStop}%
\bibitem [{\citenamefont {Ruan}\ \emph {et~al.}(2016)\citenamefont {Ruan},
  \citenamefont {Jian}, \citenamefont {Yao}, \citenamefont {Zhang},
  \citenamefont {Zhang},\ and\ \citenamefont {Xing}}]{Ruan2016}%
  \BibitemOpen
  \bibfield  {author} {\bibinfo {author} {\bibfnamefont {J.}~\bibnamefont
  {Ruan}}, \bibinfo {author} {\bibfnamefont {S.-K.}\ \bibnamefont {Jian}},
  \bibinfo {author} {\bibfnamefont {H.}~\bibnamefont {Yao}}, \bibinfo {author}
  {\bibfnamefont {H.}~\bibnamefont {Zhang}}, \bibinfo {author} {\bibfnamefont
  {S.-C.}\ \bibnamefont {Zhang}},\ and\ \bibinfo {author} {\bibfnamefont
  {D.}~\bibnamefont {Xing}},\ }\href@noop {} {\bibfield  {journal} {\bibinfo
  {journal} {Nat. Commun.}\ }\textbf {\bibinfo {volume} {7}},\ \bibinfo {pages}
  {11136} (\bibinfo {year} {2016})}\BibitemShut {NoStop}%
\bibitem [{\citenamefont {Sodemann}\ and\ \citenamefont
  {Fu}(2015)}]{Sodemann2015}%
  \BibitemOpen
  \bibfield  {author} {\bibinfo {author} {\bibfnamefont {I.}~\bibnamefont
  {Sodemann}}\ and\ \bibinfo {author} {\bibfnamefont {L.}~\bibnamefont {Fu}},\
  }\href@noop {} {\bibfield  {journal} {\bibinfo  {journal} {Phys. Rev. Lett.}\
  }\textbf {\bibinfo {volume} {115}},\ \bibinfo {pages} {216806} (\bibinfo
  {year} {2015})}\BibitemShut {NoStop}%
\bibitem [{\citenamefont {Ma}\ \emph {et~al.}(2019)\citenamefont {Ma},
  \citenamefont {Xu}, \citenamefont {Shen}, \citenamefont {MacNeill},
  \citenamefont {Fatemi}, \citenamefont {Chang}, \citenamefont {Valdivia},
  \citenamefont {Wu}, \citenamefont {Du}, \citenamefont {Hsu}, \citenamefont
  {Fang}, \citenamefont {Gibson}, \citenamefont {Watanabe}, \citenamefont
  {Taniguchi}, \citenamefont {Cava}, \citenamefont {Kaxiras}, \citenamefont
  {Lu}, \citenamefont {Lin}, \citenamefont {Fu}, \citenamefont {Gedik},\ and\
  \citenamefont {Jarillo-Herrero}}]{Ma2019}%
  \BibitemOpen
  \bibfield  {author} {\bibinfo {author} {\bibfnamefont {Q.}~\bibnamefont
  {Ma}}, \bibinfo {author} {\bibfnamefont {S.-Y.}\ \bibnamefont {Xu}}, \bibinfo
  {author} {\bibfnamefont {H.}~\bibnamefont {Shen}}, \bibinfo {author}
  {\bibfnamefont {D.}~\bibnamefont {MacNeill}}, \bibinfo {author}
  {\bibfnamefont {V.}~\bibnamefont {Fatemi}}, \bibinfo {author} {\bibfnamefont
  {T.-R.}\ \bibnamefont {Chang}}, \bibinfo {author} {\bibfnamefont {A.~M.~M.}\
  \bibnamefont {Valdivia}}, \bibinfo {author} {\bibfnamefont {S.}~\bibnamefont
  {Wu}}, \bibinfo {author} {\bibfnamefont {Z.}~\bibnamefont {Du}}, \bibinfo
  {author} {\bibfnamefont {C.-H.}\ \bibnamefont {Hsu}}, \bibinfo {author}
  {\bibfnamefont {S.}~\bibnamefont {Fang}}, \bibinfo {author} {\bibfnamefont
  {Q.~D.}\ \bibnamefont {Gibson}}, \bibinfo {author} {\bibfnamefont
  {K.}~\bibnamefont {Watanabe}}, \bibinfo {author} {\bibfnamefont
  {T.}~\bibnamefont {Taniguchi}}, \bibinfo {author} {\bibfnamefont {R.~J.}\
  \bibnamefont {Cava}}, \bibinfo {author} {\bibfnamefont {E.}~\bibnamefont
  {Kaxiras}}, \bibinfo {author} {\bibfnamefont {H.-Z.}\ \bibnamefont {Lu}},
  \bibinfo {author} {\bibfnamefont {H.}~\bibnamefont {Lin}}, \bibinfo {author}
  {\bibfnamefont {L.}~\bibnamefont {Fu}}, \bibinfo {author} {\bibfnamefont
  {N.}~\bibnamefont {Gedik}},\ and\ \bibinfo {author} {\bibfnamefont
  {P.}~\bibnamefont {Jarillo-Herrero}},\ }\href@noop {} {\bibfield  {journal}
  {\bibinfo  {journal} {Nature}\ }\textbf {\bibinfo {volume} {565}},\ \bibinfo
  {pages} {337} (\bibinfo {year} {2019})}\BibitemShut {NoStop}%
\bibitem [{\citenamefont {Zhang}\ \emph {et~al.}(2018)\citenamefont {Zhang},
  \citenamefont {Sun},\ and\ \citenamefont {Yan}}]{Zhang2018}%
  \BibitemOpen
  \bibfield  {author} {\bibinfo {author} {\bibfnamefont {Y.}~\bibnamefont
  {Zhang}}, \bibinfo {author} {\bibfnamefont {Y.}~\bibnamefont {Sun}},\ and\
  \bibinfo {author} {\bibfnamefont {B.}~\bibnamefont {Yan}},\ }\href@noop {}
  {\bibfield  {journal} {\bibinfo  {journal} {Phys. Rev. B}\ }\textbf {\bibinfo
  {volume} {97}},\ \bibinfo {pages} {041101} (\bibinfo {year}
  {2018})}\BibitemShut {NoStop}%
\bibitem [{\citenamefont {Gao}\ \emph {et~al.}(2020)\citenamefont {Gao},
  \citenamefont {Zhang},\ and\ \citenamefont {Zhang}}]{Gao2020}%
  \BibitemOpen
  \bibfield  {author} {\bibinfo {author} {\bibfnamefont {Y.}~\bibnamefont
  {Gao}}, \bibinfo {author} {\bibfnamefont {F.}~\bibnamefont {Zhang}},\ and\
  \bibinfo {author} {\bibfnamefont {W.}~\bibnamefont {Zhang}},\ }\href@noop {}
  {\bibfield  {journal} {\bibinfo  {journal} {Phys. Rev. B}\ }\textbf {\bibinfo
  {volume} {102}},\ \bibinfo {pages} {245116} (\bibinfo {year}
  {2020})}\BibitemShut {NoStop}%
\end{thebibliography}

%

\end{document}